\documentclass[]{interact}

\usepackage{epstopdf}
\usepackage[caption=false]{subfig}
\usepackage{url}
\usepackage{hyperref}
\usepackage{braket}
\usepackage{colortbl}
\usepackage[numbers,sort&compress]{natbib}
\bibpunct[, ]{[}{]}{,}{n}{,}{,}
\renewcommand\bibfont{\fontsize{10}{12}\selectfont}
\makeatletter
\def\NAT@def@citea{\def\@citea{\NAT@separator}}
\makeatother

\theoremstyle{plain}

\theoremstyle{definition}

\theoremstyle{remark}

\begin{document}

\articletype{REVIEW ARTICLE}

\title{Quantum Machine Learning: from physics to software engineering}

\author{
\name{
Alexey~Melnikov\textsuperscript{a}\thanks{CONTACT Alexey~Melnikov. Email: alexey@melnikov.info.
\begin{center}
\fbox{
\begin{minipage}{1\textwidth}
Please check the published version, which includes all the latest additions and corrections:\\
Alexey Melnikov, Mohammad Kordzanganeh, Alexander Alodjants, Ray-Kuang Lee (2023) Quantum machine learning: from physics to software engineering, Advances in Physics: X, 8:1, DOI: \href{https://doi.org/10.1080/23746149.2023.2165452}{10.1080/23746149.2023.2165452}
\end{minipage}
}
\end{center}} and Mohammad~Kordzanganeh\textsuperscript{a} and Alexander~Alodjants\textsuperscript{b} and Ray-Kuang~Lee\textsuperscript{c,d,e,f}} 
\affil{\textsuperscript{a}Terra Quantum AG, 9000 St. Gallen, Switzerland; 
\textsuperscript{b}ITMO University, 197101 St. Petersburg, Russia;
\textsuperscript{c}Institute of Photonics Technologies, National Tsing Hua University, Hsinchu 300, Taiwan;
\textsuperscript{d}Department of Physics, National Tsing Hua University, Hsinchu 300, Taiwan;
\textsuperscript{e}Physics Division, National Center for Theoretical Sciences, Taipei 10617, Taiwan;
\textsuperscript{f}Center for Quantum Technology, Hsinchu 30013, Taiwan}
}

\maketitle

\begin{abstract}
Quantum machine learning is a rapidly growing field at the intersection of quantum technology and artificial intelligence. This review provides a two-fold overview of several key approaches that can offer advancements in both the development of quantum technologies and the power of artificial intelligence. Among these approaches are quantum-enhanced algorithms, which apply quantum software engineering to classical information processing to improve keystone machine learning solutions. In this context, we explore the capability of hybrid quantum-classical neural networks to improve model generalization and increase accuracy while reducing computational resources. We also illustrate how machine learning can be used both to mitigate the effects of errors on presently available noisy intermediate-scale quantum devices, and to understand quantum advantage via an automatic study of quantum walk processes on graphs. In addition, we review how quantum hardware can be enhanced by applying machine learning to fundamental and applied physics problems as well as quantum tomography and photonics. We aim to demonstrate how concepts in physics can be translated into practical engineering of machine learning solutions using quantum software.
\end{abstract}

\begin{keywords}
Quantum information and computing, machine learning, quantum technologies, quantum and quantum-inspired algorithms, variational quantum circuits, quantum neural networks, quantum walks, graph theory, quantum tomography, photonic quantum computing.
\end{keywords}

\section{Introduction}
Nowadays due to exponential grows of information, computational speedup, acceleration of information transmission and recognition, many key global interdisciplinary problems for modern societies emerge~\cite{Noor}. In everyday life we face a problem of big data everywhere. 
Classical information science and relevant technological achievements in communication and computing  enable  to move our society  from  Internet of computers  to the Internet of Things (IoT), when humans interact with  spatially  distributed smart  systems including  high precision sensors, various  recommendation systems  based on huge amount of  on-line  information processing and its  recognition \cite{Friedemann}. Artificial intelligence  (AI) and machine learning  (ML) drive the progress in this movement of our society. These tasks and facilities require online information recognition  that is actually possible only on the basis of the parallel information processing.

\begin{figure}
    \centering
    \includegraphics[width=0.8\textwidth]{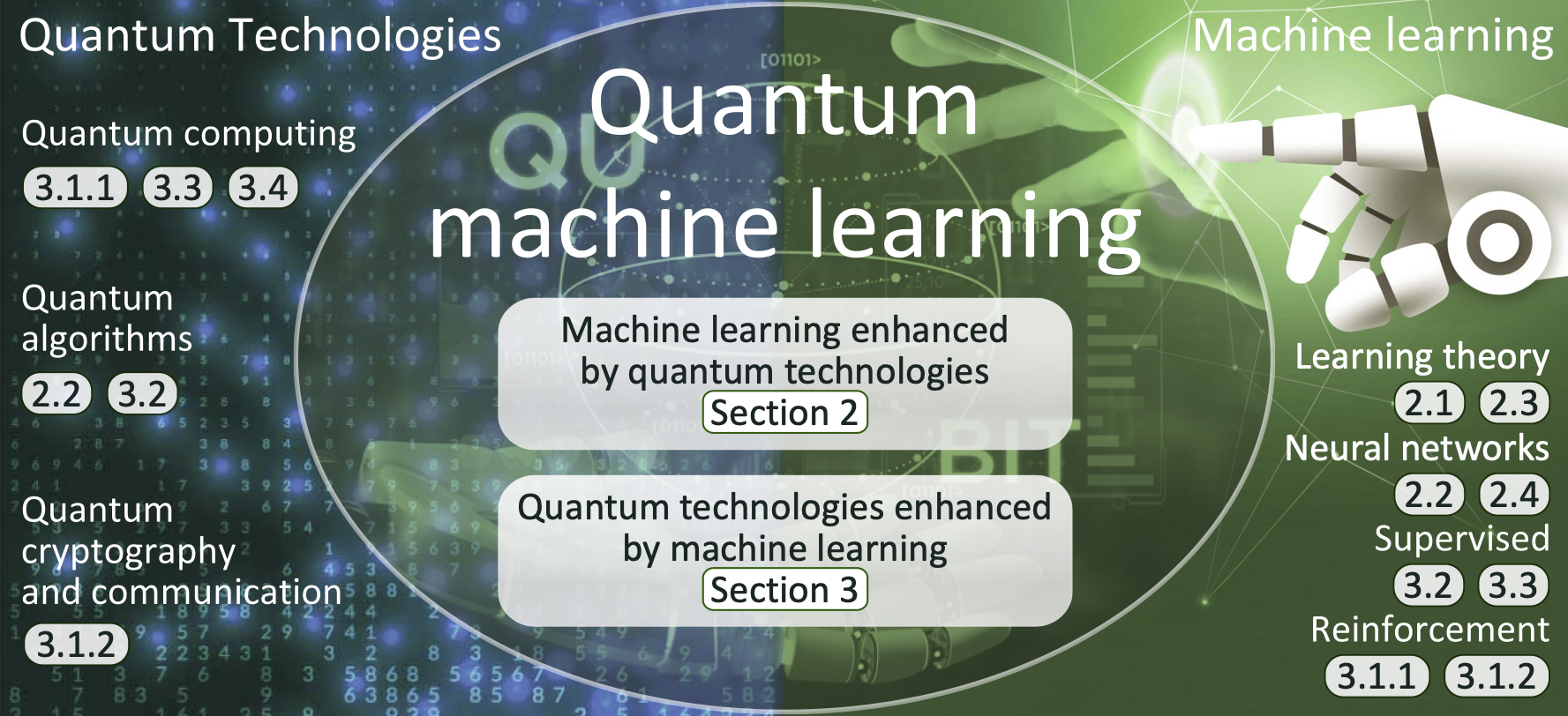}
    \caption{Interdisciplinary paradigm of quantum machine learning that is based on current classical information, quantum technologies, and artificial intelligence, respectively (the details are given in the text).}
    \label{fig:1a}
\end{figure}

Today, a number of areas have formed in information science, physics, mathematics and engineering which propose to solve these problems by means of various approaches of parallel processing of information by spatially distributed systems. Our vision of the problem we establish schematically in Fig.~\ref{fig:1a} that reflects content of this paper.  Nowadays AI predominantly focuses on ML approach that provides solutions for Big data problem, data mining problem, explainable AI, and knowledge discovery. As a result, in our everyday life we can find  distributed intelligent systems  (DIS) which represent networks of natural intelligent agents (humans) interacting with artificial intelligence agents (chatbots, digital avatars, recommendation systems,  etc. ), see e.g.~\cite{Guleva}  and references therein. Such systems require new approaches to data processing that may be described by means of cognitive computing which possess  human  cognitive capabilities, cf. \cite{Mittal}.  At the same time such a system operates within a lot of uncertainties which may provide new complexities. But, how about quantum approach and quantum technologies  which can help us in this way?

Certainly, quantum approach and relevant quantum technologies  are one of the drivers of current progress in  development of information sciences and  artificial intelligence, which have common  goals of designing efficient computing, as well as fast and secure communication and smart IoT. The mutual overlapping of the three seminal  disciplines are bearing  meaningful  fruits today. Within left halve of ellipsis  in Fig.~\ref{fig:1a} we establish some crucial topics of quantum technologies studies which are  interdisciplinary right now. In particular, quantum computing opens new horizons for classical software engineering, see e.g.~\cite{Johnston}. Especially it is necessary to mention quantum inspired algorithms and quantum inspired approaches which utilize quantum probability and quantum measurement theory for classical computing~\cite{Khrennikov}.

Quantum computers as physical systems, biological neurons and human brain are capable for parallel information proceeding in natural way. However, sufficient criterion for speedup information processing is still unknown in many cases.

A qubit which is minimal tool of quantum information science, is established by superposition of well distinguished two quantum states defined in Hilbert space and represents indispensable ingredient for parallel information processing \cite{nielsen-book}. Quantum algorithms (software) which are proposed many years ago utilize qubits quantum superposition and entanglement power for achieving so-called quantum supremacy and speedup in solution of NP-hard problems which unattainable with classical algorithms \cite{Hidary}. Quantum computers (hardware) as physical devices firstly was proposed by Richard Feynman, deals with simple two-level systems as physical qubits performing quantum computation \cite{Feynman, Manin}. Despite the fact that a lot of time has passed since the successful demonstration of the first quantum gates and the simplest operations with them (see e.g. \cite{Ladd}), there exists a large gap between the quantum information theory, quantum algorithms, and quantum computers designed to execute them. Existing quantum computers and simulators are still very far from quantum supremacy demonstrations in solving real problems related to our daily life. This can be partly explained by the modern noisy intermediate-scale quantum (NISQ) era of the development of quantum technologies \cite{Preskill}. Currently, quantum computers are restricted by small number of qubits, and relatively high level of various noises which include decoherence processes that completely (or, partially) destroy the effects of interference and entanglement. In this regard, the problem of quantum supremacy for specific tasks represents the subject of heated debates \cite{Harrow, Arute, perelshtein2022solving}. 

Surface codes and creation of logical qubits are purposed for significant reduction of computation errors \cite{Gottesman, Terhal}. 
In particular, such a codes presume mapping of some graph of physical qubits onto the logical qubit. Typically, special network-like circuits are designed for quantum processor consisting of logical qubits. However, it is unclear how this mapping is unique and how such a networks is optimal and universal for various computation physical platforms?

As an example, minor embedding procedure is supposed for quantum annealing computers which are based on superconductor quantum hardware \cite{Choi}. Obviously, various physical platforms examined now for quantum computation can use different mapping procedures and relay to design of specific networks of qubits accounting specific noises and decoherence. Thus, the choosing of appropriate network architecture represent keystone problem for current quantum computing and properly relays to demonstration of quantum supremacy. 

Clearly, the solution of this problem is connected not only with the properties of quantum systems, but also with the ability of networks to parallel and robust information processing. An important example that we refer here is a human brain as a complex network comprising from biologically active networks which exhibit fast information processing. Noteworthy, the architecture of such a computations is a non-von Neumann. In order, human brain capable for pattern retrieving by means of association.  A long time ago, Hopfield  introduced simple neural network model for associative memory \cite{Hopfield}. As time evolves, neural networks have represented an indispensable tool for parallel classical computing. Artificial intelligence and machine learning paradigm, cognitive and neuromorphic computing, use neural network some specific peculiarities represent vital approach proposed to explore the full power of parallel character of computation~\cite{Noor, Kendall}. 

\begin{figure}
    \centering
    \includegraphics[width=0.8\textwidth]{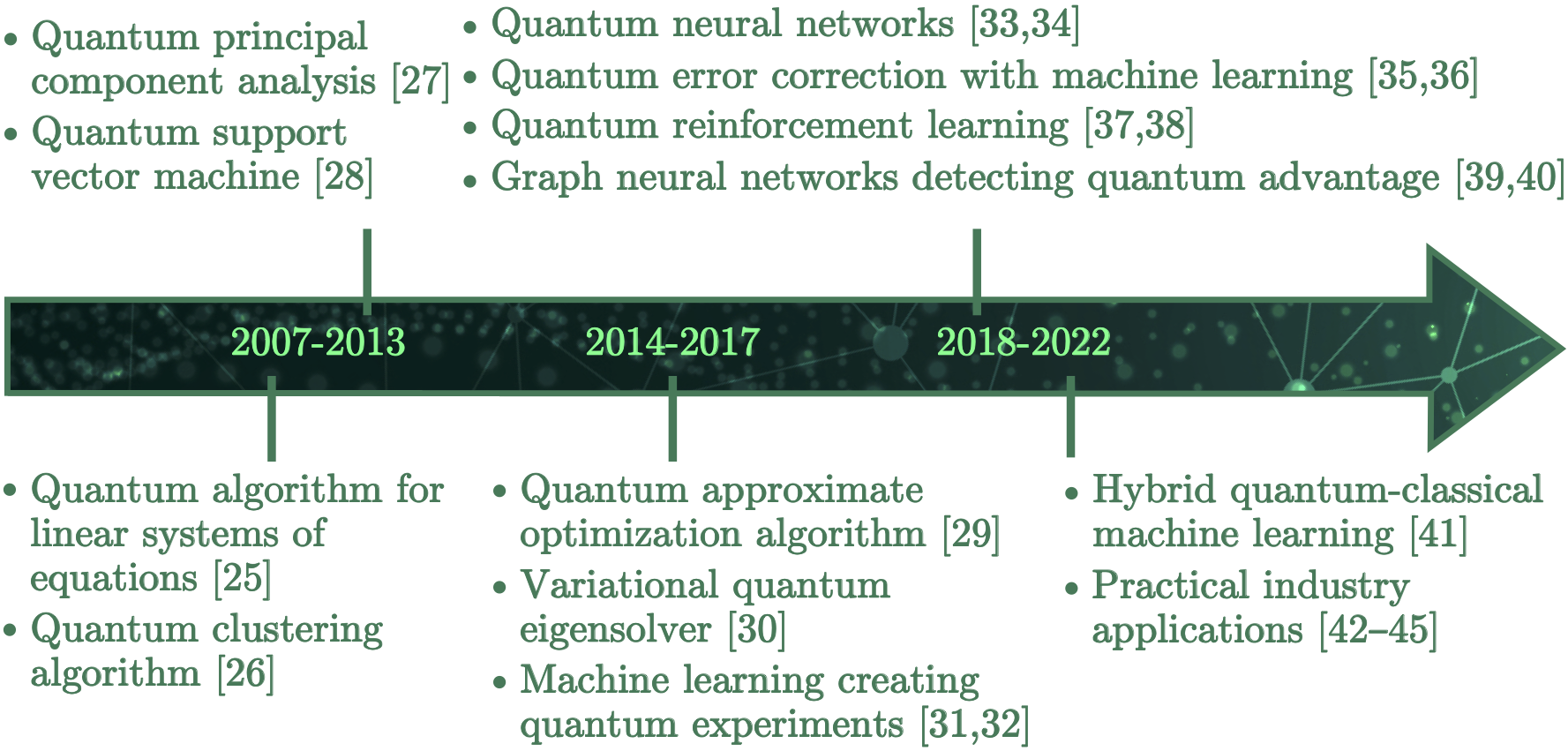}
    \caption{Timeline with milestones of quantum machine learning achievements.}
    \label{fig:1b}
\end{figure}

Quantum machine learning (QML) is a new paradigm of parallel computation where quantum computing meets network theory for improving computation speed-up and current, NISQ-era quantum technology facilities, by means of quantum or classical computational systems and algorithms \cite{Biamonte, Sinayskiy,dunjko2018machine,carleo2019machine}. 
In Fig.~\ref{fig:1b} we represent a timeline of the appearance and development of some important algorithms~\cite{hhl,BrassardQCA2007,lloydQPCA2013,lloydQSVA2013,farhi2014quantum,vqe,PhysRevLett.116.090405,melnikov2018active,farhi2019QNN,qcnn,fosel2018reinforcement,nautrup2019optimizingquantum,jerbi2021parametrized,qrl4,melnikov2019predicting,melnikov2020machinetransfer,boston-housing,q-fin-4,batra2021quantum,sagingalieva2022hyperparameter,sagingalieva2022hybrid} which are able to improve computational complexity, accuracy and efficiency within various type of hardware  available now. In this work we are going to discuss most of them  in details.

In more general, nowadays  QML  disciplines  occur at the border of current quantum technologies and artificial intelligence and includes all their methods and approaches to information processing, see Fig.\ref{fig:1a}. The rapidly growing number of publications and reviews in this discipline indicates an increasing interest in it from the scientific community, see e.g. \cite{DeBenedictis-Rev,Sajjan-Rev,Ma-Rev, Chen-Rev, Ciliberto-Rev,Zhang-Rev,Allcock-Rev,Martin-Rev}. In particular, seminal problems in algorithm theory which are capable to enhancement of quantum computing by means of ML approach we can find in \cite{Biamonte,DeBenedictis-Rev,Ciliberto-Rev,Martin-Rev}. Some applications of ML approach to solve timely  problems in material science, photonics, physical chemistry  reader can find in \cite{Sajjan-Rev,Ma-Rev, Chen-Rev, Zhang-Rev,Allcock-Rev}, respectively. 

It is important  to notice that ML approach is closely connecting with knowledge discoveries in  modern fundamental physics  which closely connected with a problem of big data and their recognition. In order, we talk about automated scientific discovery  which can significantly expand our knowledge of Nature, cf. \cite{vervoort2021artificial,Vervoort}.  In particular, it is worth to mention research of the Large Hadron Collider (LHC), where data mining can contribute to new discoveries in the field of fundamental physics \cite{Karagiorgi2022}. Another important example  constitutes  network researches on registration of gravitational waves and extreme weak signals in astronomy, see e.g. \cite{Abot}. Clearly, further  discoveries in this area require improvement of sensitivity of network detectors (which are interferometers) and obtained data mining where ML approach can significantly promote, cf. \cite{Vajente}. 

Despite the fact that previous review papers   \cite{Biamonte, Sinayskiy, DeBenedictis-Rev,Sajjan-Rev, Chen-Rev, Ciliberto-Rev,Zhang-Rev,Allcock-Rev,Martin-Rev} theoretically  substantiates and discusses the effectiveness of quantum approaches and quantum algorithms in ML problems, in practice there are many problems that do not allow  to see quantum supremacy in experiment. Within the NISQ era of modern quantum computers and simulators, their capabilities are not yet enough to achieve quantum supremacy, cf.~\cite{Preskill}.  In this regard, hybrid  information processing algorithms that take into account the sharing of quantum and classical computers come forward. Quantum-classical  variational, quantum approximate optimization (QAOA) algorithms are very useful and effective in this case, see e.g.~\cite{vqe2, LukinQAOA, Cerezo2021,deshpande2022capturing}.  In this review work we are going to discuss various approaches which are  use for QML paradigm within current NISQ-era realities. Unlike previous work
\cite{Biamonte, Sinayskiy, DeBenedictis-Rev,Sajjan-Rev,Chen-Rev, Ciliberto-Rev,Zhang-Rev,Allcock-Rev,Martin-Rev}, below  we will focus on methods and approaches of ML that can be effective, especially for hybrid  (quantum-classical) algorithms, see  Fig.~\ref{fig:1b}.

 In its most general form, current work can be divided into two large parts which we establish as Sections 2 and 3, respectively. In particular, in Sec. 2 we consider a variety of problems where the ML approach may be enhanced by means of quantum technologies, as it is presented in Fig.~\ref{fig:1a}. In general, we speak here about speed-up of data processing by quantum computers and/or quantum simulators which we can use for classical ML purposes, see Fig.~\ref{fig:2}. An important part of these studies is devoted to optimal encoding, or embedding of classical data set into the quantum device~\cite{kordzanganeh2022exponentially}, and recognition of data set from quantum state readout. We establish comprehensive  analysis of quantum neural networks (QNN) features  as a novel models in QML whose parameters are updated classically. We discussed how such a model may be used in timely hybrid quantum-classical variational algorithms.

On the contrary, in Sec. 3 we establish currently developing QML hot directions where classical ML approach can help to solve NISQ-era quantum computing and quantum technology tasks, cf. Fig.\ref{fig:1a}. In particular, it is necessary to mention automation of quantum experiments, quantum state tomography, quantum error correction, etc. where classical ML technique may be applied. Especially, we note here ML algorithms which can be useful in recognition of quantum speedup problem of random (classical, or quantum) walks performed on various graphs. The  solution of this problem proposed by us plays essential role for both of current quantum computing hardware and software development.

\section{Machine learning enhanced by quantum technologies}

\begin{figure}[t!] 
  \centering
  \includegraphics[width=0.8\textwidth]{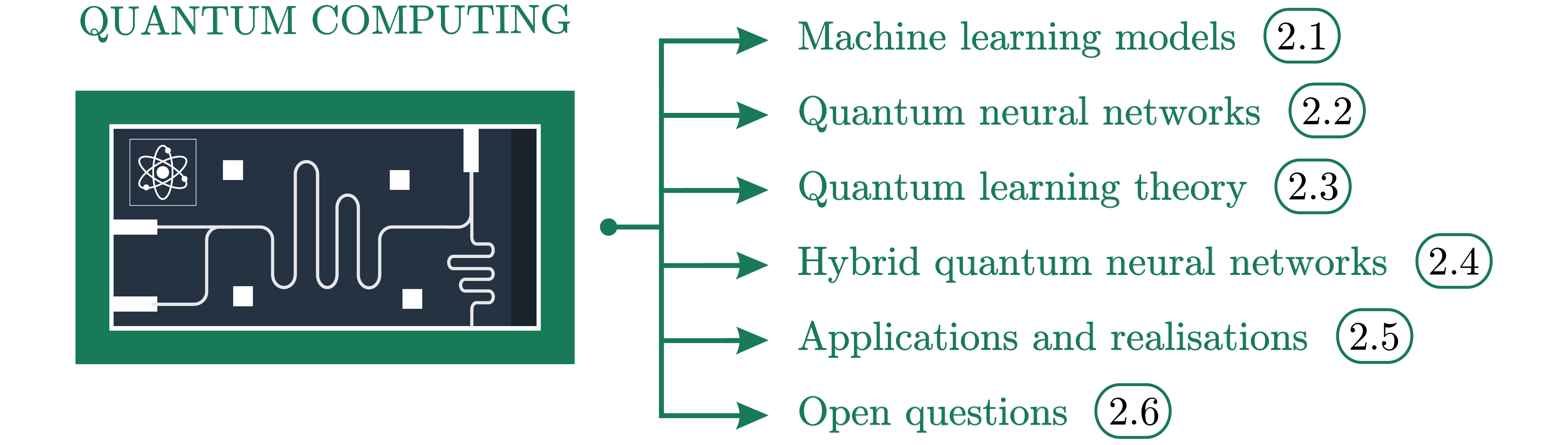}
  \caption{Quantum technologies help in improving machine learning. Sections that discuss a particular topic are labeled.}
  \label{fig:2}
\end{figure}

In this section an impact that quantum technologies make in machine learning is discussed. The outline of the topics is given in Fig.~\ref{fig:2}.

\subsection{Machine learning models}

At its roots, machine learning is a procedural algorithm that is augmented by the provision of external data to model a specific probability distribution. The data could consist of only environment variables (features), $\textbf{x} \in \mathcal{X}$ - unsupervised learning - features and their associated outcomes (labels), $\textbf{y} \in \mathcal{Y}$ - supervised learning - or environment variables and a reward for specific actions, $R(a)$ - reinforcement learning. 

\subsubsection{Unsupervised learning}

The point of unsupervised learning is to infer attributes about a series of data points, usually to find the affinity of data points to a clustering regime. A popular method of unsupervised learning is known as the K-means clustering~\cite{k-means} where data points are assigned to a chosen number of clusters, and the position of the centres of these clusters could be trained. Unsupervised learning is applied to many real-world problems, from customer segmentation in different industries~\cite{k-means-customer-segmentation} to criminal activity detection~\cite{k-means-crime}.

\subsubsection{Supervised learning}

In contrast, supervised learning endeavours to infer patterns in the provided data. The goal of such models is to generalise this inference to previously-unseen data points. In a linear regression setting, this is often done by linear interpolation \cite{linear-interpolation}, but for an a-priory problem where some degree of non-linearity is plausible, supervised learning can train non-linear regression models and provide better alternatives. Supervised learning is also used for logistical learning, where instead of a regression model, a categorical probability distribution is to be learned. Supervised learning has seen considerable success in many areas, from credit-rating models \cite{sup-credit-rating} to scientific fields \cite{sup-scientific-biology}.

\subsubsection{Reinforcement learning}

Finally, reinforcement learning is the optimisation of a set of actions (policy) in an environment. The environment allows actions and provides rewards if certain conditions are met. An agent is made to explore this environment by investigating the outcomes of certain actions given its current state and accordingly optimise its model variables. Reinforcement learning often attracts significant attention from the gaming industry \cite{rl-gaming} but it has also contributed to real-life scenarios such as portfolio management~\cite{rl-finance}.

\subsubsection{Exponential growth of practical machine learning models}

A recurring theme in all three modes of ML is the high complexity of their models. This could be caused by a high-dimensional input size like classifying a high-resolution image database \cite{classical-highD-CNN}, or a complex problem like image segmentation \cite{classical-complex-image-segmentation}. A commonly-used - but known to be inaccurate \cite{inaccurate-parameter-counting, parameter-counting-2} - measure of complexity is the parameter-count of an ML model. It is simply the number of trainable parameters of a model. Most familiarly in neural networks, the parameters are the weights and biases associated with each neural layer. For smaller problems, the parameter-count could be as small as hundreds \cite{tinyml1,tinyml2}, but cutting-edge AI models such as DALL-E 2 \cite{dalle2paper}, Gopher \cite{gopher-paper}, and GPT-3 \cite{gpt3-paper} are models that have tens or hundreds of billions of parameters and is increasing \cite{parameter-increasing-trend, parameter-increasing-trend2}. This level of high-dimensionality comes at a great financial and environmental cost. Ref. ~\cite{policy-consideration} assessed the Carbon emission and the financial cost of fine-tuning and training several large ML models in 2019. They found that in some cases these models emitted more CO2 than the entire lifetime of an average American car, and could cost over \$3m. In addition to this great cost, there are concerns regarding the scalability, namely that the exponential-growth in computing power - known as Moore's law \cite{mooreslaw}, revisited in \cite{moorerevisited} - is growing at a slower rate than ML research \cite{compute-and-ai,three-eras}. 

\subsubsection{Quantum-enhanced machine learning}

The idea behind the field of quantum machine learning is to use the capabilities of quantum computers to provide scalable machine learning models that can provide machine learning capabilities beyond what classical models can be expected to deliver, at a healthier cost. 

Quantum computers offer an exponential computational space advantage over classical computers by representing information in quantum binary digits or qubits. Where classical computers work in the Boolean space, $\mathbb{B}^{\otimes n}$, qubits form an exponentially growing, special unitary space, $\mathbb{SU}(2^n)$. This means that while a classical register with $n$ bits can hold an $n$-digit binary string, a quantum register of the same size holds all possible strings of such size, providing an exponential number of terms in comparison to its classical counterpart \cite{nielsen-book}. 

In addition to addressing the scalability concerns, classical machine learning models operate within the realm of classical statistical theory, which in some cases seems to diverge from human behavioural surveys. Ref.~\cite{sure-thing} introduced the \emph{sure thing principle}, which shows how unrelated uncertainties could affect a human's decision, which a classical statistical model would deem as unrelated and remove. In Refs.~\cite{conjunction-fallacy,sure-thing-2} it is showed that in some cases people tend to give higher credence to two events occurring in conjunction than either happening individually, which is contrary to the classical statistical theory picture. In Ref.~\cite{Khrennikov, pothos-1} it is argued that that these problems could be addressed by using a quantum statistical formulation. In addition, other similar issues like the problem of negation \cite{negation} and others listed in Ref.\cite{lewis-review} are also shown to have a resolution in the quantum theory. The distributional compositional categorical (DisCoCat) model of language \cite{discocat} could be addressed as the first theoretically successful attempt at harnessing this advantage of quantum machine learning.

\subsection{Quantum neural networks}

For a given data provision method, e.g. supervised learning, a host of different machine learning architectures could be considered. A machine learning architecture has a set of trainable parameters $\theta$ that can be realised based on an initial probability distribution. Any specific realisation of the parameters of a machine learning architecture is a model. The quest of machine learning is to train these parameters and achieve a nearer probability distribution to that of the problem in question. The fully trained version of each architecture yields a different model with different performance, and generally, the architectures that can spot and infer existent patterns in the data are said to be of superior performance. It is also important to avoid models that find non-existent patterns, models that are said to over-fit their pattern-recognition to the provided data, and when evaluated on previously-unseen data fail to perform as well. A model that can spot existent patterns without over-fitting to the provided data is said to have a high \emph{generalisation ability}. This metric establishes a platform for model selection\footnote{ Sometimes referred to as Occam's factor \cite{occam-factor}.}~\cite{mackay-book}.

For any given problem, there are a variety of architecture classes to choose from. Some of the most commonly-used architectures are multi-layered perceptrons (neural networks), convolutional networks for image processing, and graph neural networks for graphically-structured data. QML contributes to this list by introducing quantum models such as QNN~\cite{schuld-book}. 

Quantum neural networks are models in QML whose parameters are updated classically. The training pipeline includes providing data to the quantum model, calculating an objective (loss) value, and then adapting the QNN parameters in such a way as to reduce this objective function. The specific approach to providing the data to the quantum model is known as the data encoding strategy, and it can have a drastic effect on the functional representation of the model. Sec \ref{sec:encoding_strategies} covers the various approaches to data encoding, and Sec \ref{sec:fourier_estimator} offers a review of the theoretical advances in exploring the analytical form of this representation. In QNNs, the objective function is (or includes) the expectation value of a parametrised quantum circuit (PQC) \cite{pqc}. PQCs are quantum circuits that make use of continuous-variable group rotations. Fine-tuning the architecture of the PQC can have a direct effect on the performance of the resultant QNN model. Sec \ref{sec:variational_param} reviews the various PQC parametrisations suggested in the literature.

The consequences of the choice of the loss function are outlined in Sec \ref{sec:qnn_kernel}. After making this choice, one could evaluate the PQC, and pass the result to the loss function to obtain a loss value. To minimise the loss value, it is important to tune the trainable parameters in such a way to maximally minimise this value. This is achieved - in both classical and quantum ML - by calculating the gradient of the loss function with respect to the model parameters\footnote{Some alternative approaches exist known as gradient-free optimisation methods ~\cite{gradient_free_opt_methods_exist_classical}.}. The gradient vector of a function points to the direction of maximal increase in that function, and to maximally reduce the loss function one could find the gradient and step in the opposite direction. Sec \ref{sec:gradient_calculation} reviews the literature concerning QNN gradient computation. 

\subsubsection{Data encoding strategies}\label{sec:encoding_strategies}
 There are three over-arching data encoding strategies  ~\cite{schuld-kernel}:
\begin{itemize}
   \item \textbf{State embedding:} the features are mapped to the computational bases.  This is often used for categorical data, and as the number of bases grows, the number of data points needs to follow the same trend, otherwise, the encoding will be sparse~\cite{schuld-book}. 
    
    \item \textbf{Amplitude embedding:} when the features of the dataset are mapped to amplitudes of the qubit states. This embedding could be repeated to increase the complexity of the encoding. For $n$ qubits, this method allows us to encode up to $2^{n+1}$ features onto the quantum system. 
    
    \item \textbf{Observable embedding:} the features are encoded in a Hamiltonian with respect to which the quantum system is measured. This encoding is typically used in quantum native problems - see Sec \ref{sec:native_quantum} - namely variational quantum eigensolvers (VQE)~\cite{vqe} and quantum differential equation solvers~\cite{lloyd2020quantum,Kyriienko2021}.
    
\end{itemize}
It is important to recognise that state embedding is the only discrete-variable encoding with a strong resemblance to classical ML, whereas the other two are continuous-variable methods and can be considered analogue machine learning\footnote{This is subject to the input methodology. Normally, a digital computer is used to set up the quantum circuit, in which case the learning is still fully digital.}. 

Amplitude embedding could be sub-divided into sub-categories: angle embedding, state amplitude embedding, squeezing embedding, and displacement embedding \cite{embedding1,embedding2}. Ref.~\cite{schuld-kernel} provides an expressivity comparison between these encoding methods. Effective encoding strategies were analysed in~\cite{lloyd2020quantum,encoding2,encoding3,Schuld_fourier}.

\subsubsection{Parametrised architecture}\label{sec:variational_param}

The specific parametrisation of the network could dramatically change the output of a circuit. In classical neural networks, adding parameters to a network improves the model expressivity, whereas, in a quantum circuit, parameters could become redundant in over-parametrised circuits \cite{rank-paper}. Additionally, the architecture must be trainable, whereas it was shown that this cannot be assumed in an a-priori setting \cite{mcclean_2018_barren} - see Sec \ref{sec:barren-plateaus}. Many architectures have been suggested in the literature, and many templates are readily available to choose from on QML packages~\cite{PL-templates,PL-tfq-templates,PL-ibmq-templates}.

Ref.~\cite{hardware-efficient} introduced a family of hardware-efficient architectures and used them as variational eigensolvers - see Sec \ref{sec:native_quantum}. These architectures repeated variational gates and used CNOT gates to create highly-entangled systems. Based on the discrete model in Ref.~\cite{iqp-first} and made continuous in Ref.~\cite{iqp-cv} a model was devised using RZZ quantum gates that was shown to be computationally expensive to classically simulate \cite{iqp-hard,IQPpaper}, named the instantaneous polynomial-time quantum ansatz (IQP). 

Another approach to creating quantum circuits is to take inspiration from tensor networks~\cite{tensor-networks}. Famous architectures in this class are the tensor-tree network (TTN), matrix product state (MPS), and the quantum convolutional neural networks (QCNN) \cite{ttn,mps1,mps2,qcnn}.

\subsubsection{Gradient calculation}\label{sec:gradient_calculation}
Despite its excessive memory usage~\cite{backprop-memory}, the most prominent gradient calculation method in classical ML is the back-propagation algorithm \cite{backprop}. This method computes the gradient of every function that trainable parameters are passed through alongside its output and employs the chain rule to create an automatically differentiable ML routine. The back-propagation method can (and has been~\cite{pl-backprop}) implemented for QML, but as it requires access to the quantum state-vector, it can only be used on a simulator and not a real quantum processing unit (QPU). As quantum advantage can only occur in the latter setting, it is important to seek alternatives that can operate on QPUs. 

The first proposed algorithm is known as the finite-difference differentiation method \cite{finite_difference}. As its name suggests, it calculates the gradient by using the first principles of taking derivatives, i.e. adding a finite difference to the trainable parameters one at a time, and observing the change that this action makes. This method is prone to error in the NISQ era.

As an alternative, a discovery was made in~\cite{parameter-shift-paper} known as the parameter-shift rule that suggested an exact, analytic derivative could be calculated by evaluating the circuit twice for each trainable parameter. The suggestion was that the derivative of a circuit with respect to a trainable parameter $\theta$ is half of the evaluation of the circuit with $\theta$ shifted by $\frac{\pi}{2}$ subtracted from the circuit when it is shifted by $-\frac{\pi}{2}$. This suggestion initially worked only on trainable parameters applied to Pauli rotations, but later works \cite{param-shift-improv-1,param-shift-improv-2,param-shift-improv-3,param-shift-improv-4,param-shift-improv-5,param-shift-improv-6,param-shift-improv-7,param-shift-improv-8} expanded to its current form, applicable to any parameterisation. The parameter-shift rule is the state-of-the-art gradient computation method and is compatible with QPUs, but, one of its major problems is its scalability. As mentioned, the number of circuit evaluations for this method increases linearly with the number of trainable parameters, and this poses a challenge to how complex the quantum models can get. A notable effort to mitigate this effect was by parallelising the gradient computation, which is now natively provided when using PennyLane on AWS Braket \cite{PL-parallel}.

As a transitional gradient computation method for QNNs, Ref.~\cite{adjoint} introduced the adjoint algorithm. Similar to the back-propagation method, adjoint can only be run on a simulator and calculates the entire gradient vector using a single evaluation of the circuit. However, its memory usage is superior to the former. It works by holding a copy of the quantum state and its adjoint in the memory, and in turn applying the gates in reverse order, calculating the gradients wherever possible. This means that two overall evaluations of the circuit are made, first to evaluate the output, and second to compute the gradient.

Alternative suggestions have also been made to optimise QML models following the geometry of their group space. Ref.~\cite{gradient-flow} suggested a Riemannian gradient flow over the Hilbert space, which through hardware implementation showed its favourable optimisation performance. 

\subsubsection{Quantum neural networks as universal Fourier estimators}\label{sec:fourier_estimator}

Ref.~\cite{Schuld_fourier} explored the effects of data encoding on the expressivity of the model. It proved that the data re-uploading technique suggested by Ref.~\cite{data_reuploading} created a truncated Fourier series limited by the number of repetitions of the encoding gates. Ref.~\cite{qnn_universality} also showed that QNNs can be universal Fourier estimators - an analogue to the universality theorem in classical multi-layered perceptrons \cite{MLP_universality}. Another point proven by Refs.~\cite{Schuld_fourier,data_reuploading} was that by repeating the encoding strategy (in amplitude embedding and more specifically the angle embedding) more Fourier bases are added to the final functional representation of the circuit. This was true if the repetitions were added in parallel qubits or series. This sparked a question about the accessibility of these Fourier bases, i.e. whether their coefficients can independently be altered which remains an open question at the time of this publication. 

\subsubsection{Barren plateaus and trainability issues}\label{sec:barren-plateaus}

QNNs could suffer from the problem of vanishing gradients. This is when during training, the gradient of the model tends to zero in all directions. This could severely affect the efficiency of the training or even bring it to a halt. This is known as the barren plateau (BP) problem.

\begin{figure}
    \centering
    \includegraphics[width=0.8\textwidth]{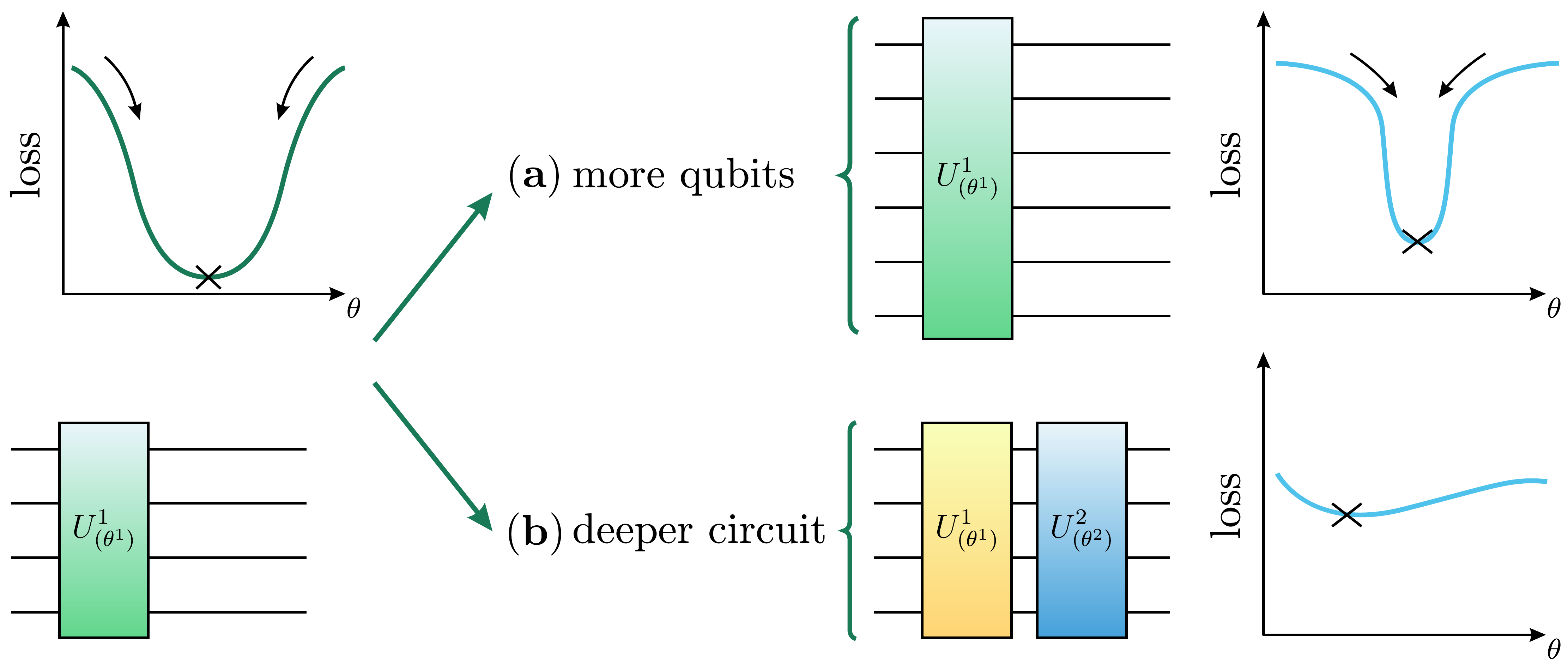}
    \caption{Visualisation of the barren plateau phenomenon in (a) noise-free and (b) noise-induced settings. }
    \label{fig:bp}
\end{figure}

BPs are not usually at the centre of attention in classical ML, but their dominance in quantum architectures makes them one of the most important factors in choosing a circuit. Ref.~\cite{mcclean_2018_barren} showed that the expectation value of the derivative of a well-parametrised\footnote{Formally when the variational architecture of a quantum circuit nears a 1-design\cite{t-design}.} quantum circuit is equal to zero, and that its variance decays exponentially with the number of qubits. Ref.~\cite{gradient-free} confirmed that barren plateaus also exist in gradient-free optimisation methods. In addition, Ref.~\cite{noise-bp} showed that in the NISQ era, using deep circuits flattens the overall loss landscape resulting in noise-induced BPs.These are mathematically different kinds of Barren plateaus that flatten the landscape as a whole. The illustrations in Fig \ref{fig:bp} summarise these phenomena.

These two findings painted a sobering picture for the future of QNNs, namely that they need to be shallow and low on qubit-count to be trainable, which contradicts the vision of high-dimensional, million-qubit QML models. 

Many remedies have been proposed: Ref.~\cite{layerwise} suggested that instead of training all parameters at once, the training could be done layer-wise and Ref.~\cite{log-nobp} showed that if the depth of the variational layers of the QNN is of the order O($\log(n)$), $n$ being the number of qubits, and that only local measurements are made, the QNN remains trainable. This was tested on circuits with up to 100 qubits and no BPs were detected. Other remedies included introducing correlations in the trainable parameters \cite{correlation-bp1,correlation-bp2} and specific initialisations of the parameters \cite{initialisation} of the circuit by applying adjoint operators of certain variational gates. 

More analysis was done on specific architectures: Ref.~\cite{ttn-nobp} showed that under well-defined weak conditions, the TTN architecture was free from the BP problem; and Ref.~\cite{qcnn-nobp} showed that the quantum convolutional neural network architecture introduced in Ref.~\cite{qcnn} was also free from BPs. Ref.~\cite{bp-zx} developed a platform based on ZX calculus~\footnote{ZX calculus is a graphical language in quantum information theory~\cite{zx1,zx2,zx3}.} to analyse whether a given QNN is subject to suffering from BPs. In addition to confirming the results from the two earlier contributions, it also proved that the matrix product state~\cite{mps-paper} and hardware efficient, strongly-entangling QNNs suffered from BPs. Furthermore, Ref.~\cite{bp-entanglement} related the barren plateau phenomenon to the degree of entanglement present in the system.

\subsection{Quantum learning theory}\label{sec:learning-theory}
\subsubsection{Supervised QML methods are kernel methods}\label{sec:qnn_kernel}
In Refs.~\cite{schuld-kernel, schuld-feature} the similarities between the QNNs and kernel models were brought to focus. First introduced in Ref.~\cite{vapnik-first}, kernel methods are well-established ML techniques with many applications. In conjunction with support vector machines (SVM), the way they work is by mapping the features of a dataset into a high-dimensional space through a function $\phi (\textbf{x})$ and then using a kernel function, $\mathcal{K}(x_1,x_2)$, as a distance measure between any two data points in this high-dimensional space. This is exactly the behaviour observed in QNNs: the features are first embedded into a high-dimensional quantum state-vector, and by overlapping one encoded state with another we can find the level of similarity between two points in this space. In this high-dimensional space, one hopes to find better insight into the data - usually expressed as a decision boundary in the form of a hyperplane in classification tasks. Ref.~\cite{power-of-data} used this link and developed a framework for searching for possible quantum advantages over classical models. It also showed that large models could scatter the data so far apart that a distance measure becomes too large for optimisation purposes, and proposed that an intermediate step be added to map the high-dimensional space into a lower-dimensional hyperplane to improve its training performance.

\subsubsection{Bayesian inference}

Bayesian inference is an alternative approach to the statistical learning theory where Bayes' theorem \cite{bayes} is used to adapt an initial assumption about the problem (prior distribution) based on newly-found data (evidence) to get a posterior distribution. Bayesian learning is when this logic is applied to ML. This is done by applying a distribution to every parameter in the network and updating the distributions when training. Calculating the posterior distribution is generally computationally expensive, but it is possible to approximate it using a trick known as variational inference~\cite{variational-inference, bayes-by-backprop} successfully demonstrated an approximate back-propagation algorithm on a Bayesian neural network (BNN), referred to as Bayes-by-backprop. 

The first implementations of Bayesian QNNs were in Refs.~\cite{qbayes1,qbayes2,qbayes3} which attempted to make quantum circuits into an exact Bayesian inference machine. Ref.~\cite{fiorentini-paper} introduced two efficient, but approximate methods - one from kernel theory and another using a classical adversary - to use QNNs to perform variational inference. The work consists of a quantum circuit that can be modelled to produce the probability distribution of a phenomenon by exploiting the probabilistic nature of quantum mechanics - known as a Born machine \cite{born-machines,q-native-born} or quantum generative models \cite{qgen1,mps-paper,qgen3,qgen4,qgen5,qgen6,qgen7}. This could also be used later to quantify the prediction error for a single data point, as it has been done classically in Ref.~\cite{kendallgal}.

\subsubsection{Model complexity and generalisation error bounds}\label{sec:complexity}
Intuitively, complex phenomena require complex modelling, but quantifying the complexity of a given model is non-trivial. There are multiple ways of defining the model complexity: Vapnik-Chervonenkis (VC) dimension~\cite{vc}, Rademacher complexity~\cite{rademacher}, and effective dimension~\cite{eff-dim-paper}\footnote{For an exhaustive list of complexity measures see \cite{parameter-counting-2}}. The complexity measures are also connected to the generalisation error because when the model becomes too complex for the problem, the generalisation is expected to worsen. 

Much work has been done to quantify the complexity and the generalisation error of quantum neural networks: Ref.~\cite{fourier-complexity} explored a generalisation error bound through the Rademacher complexity that explicitly accounted for the encoding strategy; and Ref.~\cite{amirapaper} used the effective dimension - a measure dependent on the sample size - to bound the generalisation error of QNNs as well as prove their higher expressivity given the same number of trainable parameters. Other attempts were also made to quantify the complexity (also referred to as the expressivity) of QNNs in Refs.~\cite{express1,express2,express3,express4,express5}\footnote{It is noteworthy that Ref.~\cite{peters2022generalization} showed that some models have a higher generalisation ability despite overfitting.}. Notably, Ref.~\cite{caro-generalisation} theoretically proved that the generalisation error of a QNN grew as $\mathcal{O}(\sqrt{T/N})$ where $T$ was the number of parametrised quantum gates in the QNN, and $N$ was the number of data samples in the dataset. The latter work implies that QML models are better at generalising from fewer data points.

\subsection{Hybrid quantum neural networks}\label{sec:hybrid}

\begin{figure}[t]
    \centering
    \includegraphics[width=0.8\textwidth]{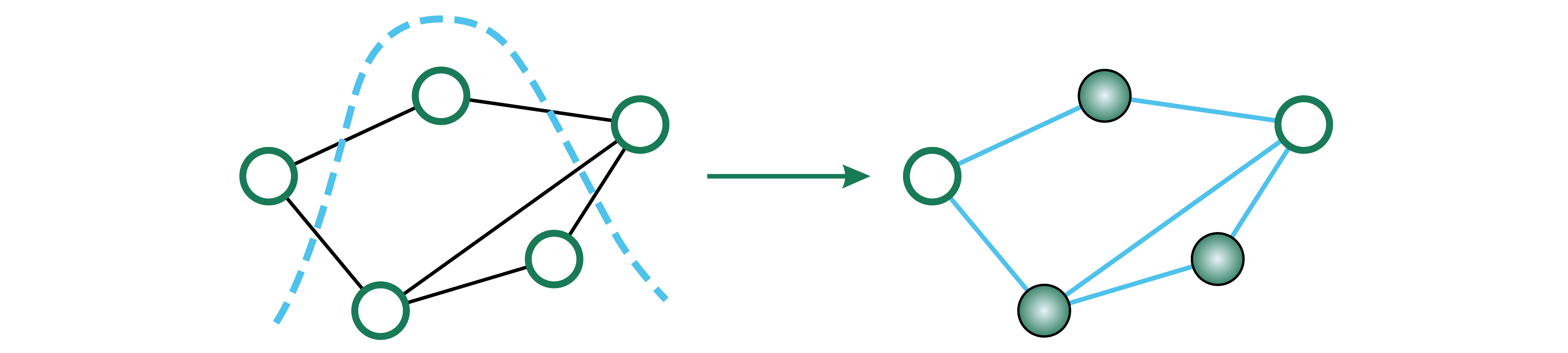}
    \caption{An example of a hybrid quantum-classical ML model.  In this case, the inputs are passed into a fully connected classical multi-layered perceptron, and its outputs are fed into the embedding of a quantum circuit. Depending on the setting, some measurements of this quantum circuit are taken and then passed into another fully-connected layer, the output of which can be compared with the label.}
    \label{fig:hybrid}
\end{figure}

Just as there are classical and quantum models, one could also combine the two to create hybrid models -- see Fig.~\ref{fig:hybrid}. It is conceivable that in the NISQ era, one could use the understanding of QML described in Sec.~\ref{sec:learning-theory} to find a regime where quantum models cover some bases that classical models do not. Ref.~\cite{boston-housing} developed a platform for hybrid quantum high-performance computing (HQC) cloud and it was deployed on the QMware hardware~\cite{qmware,kordzanganeh2022benchmarking}. It showed that for high-dimensional data, a combination of classical and quantum networks in conjunction could offer two advantages: computational speed and the quality of the solution. The data points were, first, fed to a shallow\footnote{Apart from performance considerations, this also provided an effective way to avoid barren plateaus described in Sec \ref{sec:barren-plateaus}.} quantum circuit composed of 4 qubits, two of which were measured and their corresponding values were passed onto a neural network. Two classical datasets were chosen to explore the effectiveness of a hybrid solution and to compare it when the quantum part is removed, leaving only a classical network: the sci-kit circles' dataset~\cite{circles-dataset} and the Boston housing dataset \cite{boston-dataset}. The former is a synthetic geometrical dataset that consists of two concentric circles in a 2-dimensional square of side $x = 2\pi$, and the latter is about the distribution of the property value given its population status and the number of rooms. In both cases, it was shown that the hybrid network generalises better than the classical and this difference is most visible at the extremes of problems with very small training sample sizes. However, this difference became smaller as the number of samples grew.

In continuation, Ref.~\cite{sagingalieva2022hyperparameter} suggested a hyperparameter optimisation scheme aimed at architecture selection of hybrid networks. This work also implemented a hybrid network for training, but in two new ways: 1) using a real-world, image recognition dataset~\cite{vw-dataset}, and 2) the quantum part of the hybrid network was inserted in the middle of the classical implementation. Additionally, the architecture of the quantum part was a subject of hyper-parameter optimisation, namely the number of qubits used and the number of repetition layers included were optimised. Training this network showed that the hybrid network was able to achieve better quality solutions albeit by a small margin. It is notable that because of this architecture optimisation, a highly improved quantum circuit was achieved. The performance of this circuit was theoretically measured by applying analysis methods such as ZX reducibility~\cite{zx1}, Fourier analysis~\cite{Schuld_fourier}, Fisher information~\cite{q_fisher_1,q_fisher_2,rank-paper}, and the effective dimension~\cite{eff-dim-paper,amirapaper}.

\subsection{Applications and realisations}
QML automatically inherits all classical ML problems and implementations, as it is simply a different model to apply to data science challenges. In addition to this inheritance, QML research has also provided novel, quantum-native solutions. In both cases, QML has so far been unable to provide a definite, practical advantage over classical alternatives, and all the suggested advantages are purely theoretical.

\subsubsection{Solving classical problems}

QML is employed in many classical applications. Some notable contributions are in sciences~\cite{q-sci1,q-sci2,q-sci3,q-sci4,q-sci5,q-sci6,Guan_2021}, in finance~\cite{q-fin-1,q-fin-2,q-fin-3,q-fin-4}, pharmaceutical~\cite{batra2021quantum,sagingalieva2022hybrid}, and automotive industries~\cite{sagingalieva2022hyperparameter}. In many cases, these models replaced a previously-known classical setting \cite{qmodel1,qmodel2,qmodel3,qmodel4}. Quantum generative adversarial networks were suggested in Ref.~\cite{qGAN} and followed by Refs.~\cite{qGAN1,qGAN2,qGAN3,qGAN4,qGAN5,qGAN6,qGAN7,qGAN8}. Similarly, quantum recurrent neural networks were investigated in \cite{qrnn,qrnn2} and two approaches to image recognition were proposed in Refs.~\cite{qcnn,quanvolutional}. Ref.~\cite{q-nlp} looked at a classical-style approach to quantum natural language processing. The applications of QML in reinforcement learning were also explored in Refs.~\cite{qrl1,qrl2,qrl3,qrl4}\footnote{Notably the first photonic continuous-variable policy network was introduced in Ref.~\cite{nagy2021photonic}.}. Finally, a celebrated application is the quantum auto-encoder, where data is compressed and then re-constructed from less information, a notable suggestion was made in Ref.~\cite{QAE}.

\subsubsection{Quantum-native problems} \label{sec:native_quantum}

Native problems are novel, quantum-inspired ML problems that are specifically designed to be solved by a QML algorithm.

Perhaps the most known QML algorithm is the variational quantum eigensolver (VQE). The problem formulation is that the input data is a Hamiltonian and we are required to find its ground state and ground-state energy. The VQE solution consists of preparing a PQC of trainable parameters and taking the expectation value of the Hamiltonian. This yields the energy expectation of the prepared state, and the idea is that by minimising this expectation value, we can achieve the ground-state energy, at which point the prepared state will represent the ground state of the problem. This was first implemented to find the ground-state energy of $\text{He}-\text{H}^+$~\cite{vqe} and was then substantially extended in Ref.~\cite{vqe2}. VQE remains one of the most promising areas of QML. 

Ref.~\cite{hhl} showed that PQCs can be used to solve a linear system of equations (LSE). They proposed a commonly known as the Harrow, Hassidim, and Lloyd (HHL). Refs.~\cite{de1,de2,de3,de4,de5,de6,de7,de8} improved this algorithm and Ref.~\cite{lloyddepaper} extended it to also include non-linear differential equations. Ref.\cite{Kyriienko2021} showed that it is possible to use a quantum feature map to solve non-linear differential equations on QNNs. This is also an exciting and promising area of QML. 

\begin{figure}
    \centering
    \includegraphics[width=0.8\textwidth]{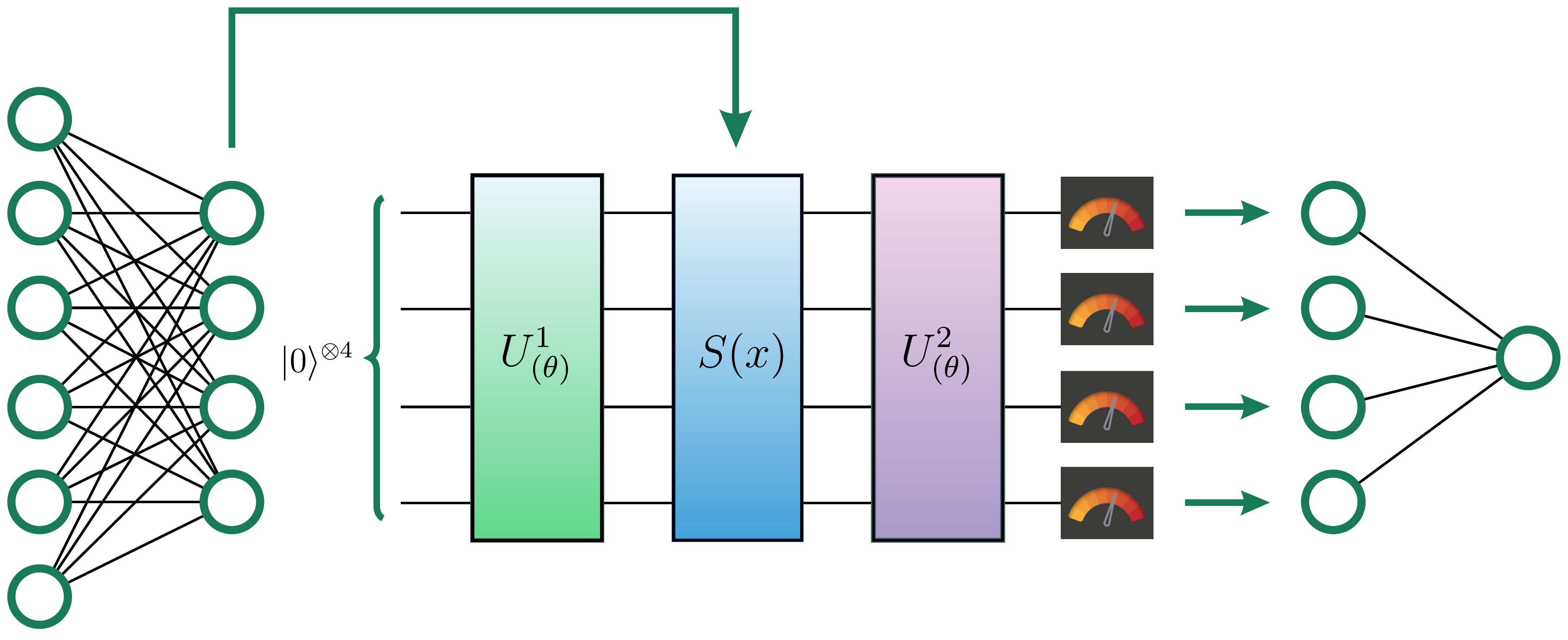}
    \caption{The max cut problem. The abstract manifestation of this problem is a general graph, and we are interested in finding a partition of its vertices such that the number of edges connecting the resultant graph to the complimentary graph is maximal.}
    \label{fig:max-cut}
\end{figure}

An important QML formulation is known as the quadratic unconstrained binary optimisation (QUBO) \cite{qubo1,qubo2}. This is generally a classical problem, but using the Ising model - see \cite{ising} - this can be solved on a quantum computer~\cite{qubo-quantum,qubo-quantum-2}. A common demonstration of the latter is the max cut problem~\cite{max-cut} - see Fig \ref{fig:max-cut}. There are solutions for the QUBO problem on both gate-based quantum computers and quantum annealers \cite{annealer1,annealer2,annealer3}, and this general concept has seen use in many sectors \cite{qubo-app}. 

Lastly, another quantum-native formulation is in natural language processing. Ref.~\cite{discocat} developed a platform for turning grammatical sentences into quantum categories using J. Lambek's work~\cite{lambek}. Refs.~\cite{qnlp1,qnlp2,qnlp3} tested this algorithm on quantum hardware, and later a full QNLP package was developed~\cite{lambeq}. The initial value proposition of QNLP in this way is that this algorithm is natively grammar-aware, but given that large classical language models are shown to infer grammar~\cite{gpt3-paper}, the real advantage of this approach could lie in other avenues, such as a potential Grover-style~\cite{grover-paper} speed-up in text classification.

\subsection{Open questions}
\subsubsection{Quantum advantage}
Despite the theoretical findings in Sec \ref{sec:complexity}, there is limited demonstrable success in using QML in real-life problems, and this is not purely due to hardware shortcomings. Ref.~\cite{rig-robust} showed that there exists a class of datasets that could showcase quantum advantage, and Ref.~\cite{power-of-data} found a mathematical formulation for where we can expect to find such an advantage. In Refs.~\cite{q_adv_1,q_adv_2,q_adv_3} attempts were also made to devise a set of rules for potential quantum advantage. However, Ref.~\cite{schuld-advantage} argued that a shift of perspective from quantum advantage to alternative research questions could unlock a better understanding of QML. The suggested research questions were: finding an efficient building block for QML, finding bridges between statistical learning theory and quantum computing, and making QML software ready for scalable ML\footnote{Other examples of include quantum federated (distributed) learning (QFL)~\cite{chehimi2022quantum}. Notably for the latter, Ref.~\cite{sheng2017distributed} proposed the first model of distributed secure quantum machine learning}.

\subsubsection{Optimal parametrisation}
In Sec \ref{sec:variational_param}, we encountered various QNN parametrisations with specific properties. An open question is how to optimally parametrise a circuit to avoid barren plateaus, be as expressive as possible, and be free of redundancy. A potential characteristic of such parametrisation is a high level of Fourier accessibility as mentioned in Sec \ref{sec:fourier_estimator}, potentially requiring a quantifiable measure of this accessibility.

\subsubsection{Theory for hybrid models}

Despite the successes outlined in Sec \ref{sec:hybrid}, the theoretical grounding for such models is limited. We saw that hybrid networks performed well if the quantum section was introduced at the beginning of the model architecture~\cite{boston-housing} or in the middle~\cite{sagingalieva2022hyperparameter}. From an information-theoretic perspective, this needs to be investigated in more detail to shed light on the effect of hybridisation. Such investigation could identify if there exist areas where the application of a quantum part could complement a classical circuit by either introducing an information bottle-neck to prevent over-fitting or by creating high-dimensional models.

\subsubsection{An efficient optimisation method}

The current gradient calculation methods are either only available on simulators or require a linearly-growing number of circuit evaluations - see Sec.~\ref{sec:gradient_calculation}. Neither of these can accommodate a billion-parameter, million-qubit setting. This poses a barrier to the future of QML, and thus an efficient optimisation method is needed for the long term. 

\section{Quantum technologies enhanced by machine learning}

\begin{figure}[t!] 
  \centering
  \includegraphics[width=0.8\textwidth]{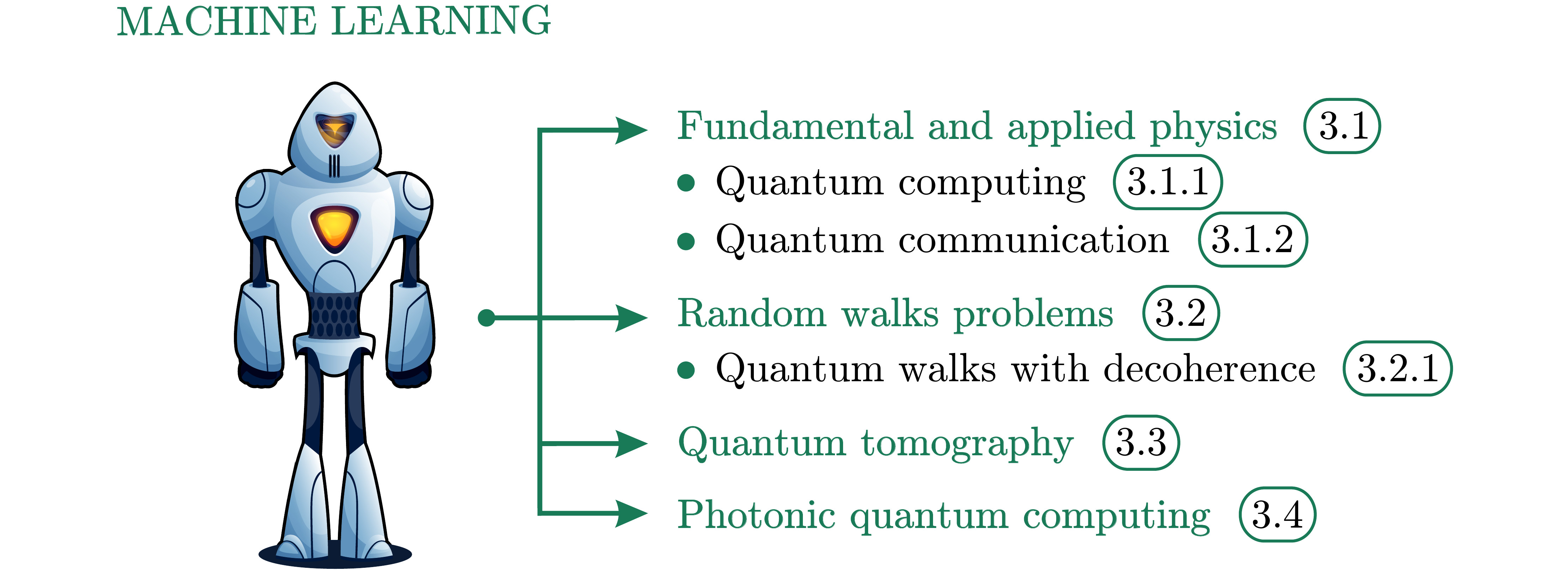}
  \caption{Machine learning helps in solving problems in fundamental and applied quantum physics. Sections that discuss a particular problem are labeled.}
  \label{fig:6}
\end{figure}

In this section an impact that machine learning makes in quantum technologies is discussed. The outline of the topics is given in Fig.~\ref{fig:6}. Today machine learning is used to realize algorithms and protocols in quantum devices, by autonomously learning how to control~\cite{fosel2018reinforcement,bukov2018reinforcement,xu2019generalizable,luchnikov2020machine,schafer2020differentiable}, error-correct~\cite{tiersch2015adaptive,nautrup2019optimizingquantum,sweke2020reinforcement,agnes2019hamiltonian}, and measure quantum devices~\cite{liu2020repetitive}. Given experimental data, ML can reconstruct quantum states of physical systems~\cite{shang2019reconstruction,torlai2019integrating,palmieri2020experimental,ding2020retrieving}, learn compact representations of these states~\cite{carleo2017solving,gao2017efficient}, and validate the experiment~\cite{agresti2019pattern}. In this section we discuss the impact of machine learning on fundamental and applied physics, and give specific examples from quantum computing and quantum communication.

\subsection{Machine learning in fundamental and applied quantum physics}

Since its full development in the mid-1920s, a century later quantum mechanics is still considered as the most powerful theory, modeling a wide range of physical phenomena from subatomic to cosmological scales with the most precise accuracy. Even though the  measurement problem and quantum gravity had led many physicists to conclude that quantum mechanics cannot be a complete theory, the spooky action of entanglement in the  Einstein-Podolsky-Rosen pair~\cite{PhysRev.47.777}, has provided the resources for quantum information processing tasks. With machine learning, one may be able to model different physical systems (e.g., quantum, statistical, gravitational) using artificial neural networks, which might lead to the development of a new framework for fundamental physics.

Even without a precise description of a physical apparatus and solely based on measurement data, one can prove the quantumness of some observed correlations by the device-independent test of Bell nonlocality~\cite{Sekatski2021deviceindependent}. In particular, by using generative algorithms to blend automatically many multilayer perceptrons (MLPs), a machine learning approach may allow the detection and quantification of nonlocality as well as its quantum (or postquantum) nature~\cite{canabarro2019machine, melnikov2020setting,valcarce2022automated}.

\subsubsection{Machine learning in quantum computing}

Machine learning has also became an essential element in applied quantum information science and quantum technologies. ML that was inspired by the success of automated designs~\cite{PhysRevLett.116.090405}, was demonstrated to be capable of designing new quantum experiments~\cite{melnikov2018active}. 

\begin{figure}[t!] 
  \centering
  \includegraphics[width=0.8\textwidth]{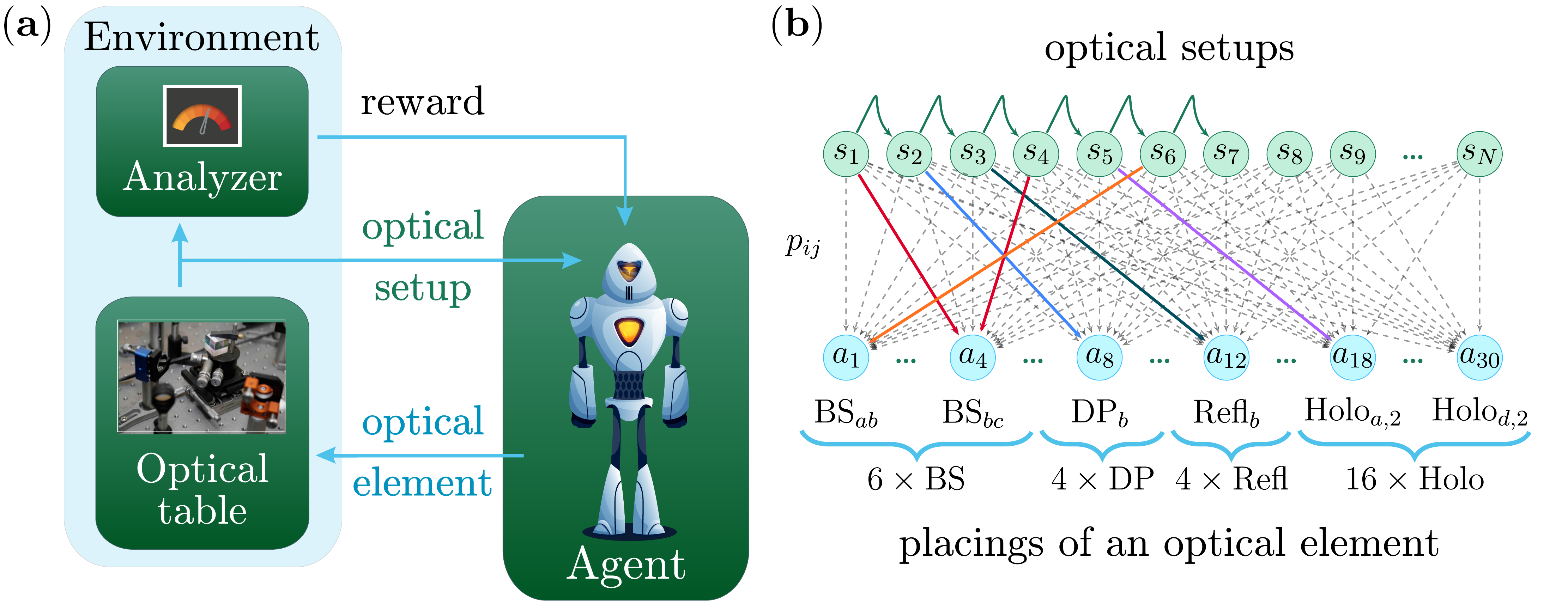}
  \caption{A reinforcement learning algorithm that designs a quantum experiment. An experiment on an optical table is shown as an example. (a) The learning scheme depicts how an agent, the reinforcement learning algorithm, learns to design quantum experiments. (b) Representation of the reinforcement learning algorithm, projective simulation, as a two-layered network of clips.}
  \label{fig:MLforQExp}
\end{figure}

Quantum experiments represent an essential step towards creating a quantum computer. More specifically, for example, three-particle photonic quantum states represent a building block for a photonic quantum computing architecture. In Ref.~\cite{melnikov2018active} ML algorithm used is a reinforcement learning algorithm based on the projective simulation model~\cite{briegel2012projective,mautner2013projective,makmal2016meta,melnikov2017projective,melnikov2018benchmarking,PScode}. An agent, the reinforcement learning algorithm, puts optical elements on a (simulated) optical table. Each action adds an optical element to an existing setup. In case the resulting setup achieves the goal, e.g., creates a desired multiphoton entangled state, the agent receives a reward. The described learning scheme is depicted in Fig.~\ref{fig:MLforQExp}(a).

The initial photonic setup is an output of a double spontaneous parametric down-conversion (SPDC) process in two nonlinear crystals. Neglecting these higher-order terms in the down-conversion, the initial state $\ket{\psi(0)}$ can be written as a tensor product of two orbital angular momentum entangled photons,
\begin{equation}
	\ket{\psi_0} = \frac{1}{3}\left(\sum_{m=-1}^{1} \ket{m}_a\ket{-m}_b\right) \otimes \left(\sum_{m=-1}^{1} \ket{m}_c\ket{-m}_d\right),
\label{eq:initial}
\end{equation}
where the indices $a, b, c$ and $d$ specify four arms in the optical setup. The actions available to the agent consist of beam splitters (BS), mirrors (Refl), shift-parametrized holograms (Holo), and Dove prisms (DP). The final photonic state $\ket{\psi_f}$ is obtained by measuring the arm $a$, and post-selecting the state in the other arms based on the measurement outcome in $a$.

The reinforcement learning algorithm that achieves the experimental designs is shown in Fig.~\ref{fig:MLforQExp}(b). It is the projective simulation agent represented by a two-layered network of clips. The first layer corresponds to the current state of the optical setup, whereas the second layer is the layer of actions. The connectives between layers define the memory of the agent, which changes during the learning process. The connectivities correspond to the probabilities of reaching a certain action in a given state of a quantum optical setup. During the learning process the agent automatically adjusts the connectivities, and thereby prioterize some actions other the other. As shown in Ref.~\cite{melnikov2018active} this leads to a variety of entangled states of improved efficiency of their realization.

\subsubsection{Machine learning in quantum communication}

In addition to designing new experiments, ML helps in designing new quantum algorithms~\cite{cincio2018learning} and protocols~\cite{wallnofer2020machine}. Designing new algorithms and protocols has similarities to experiment design. In particular, similar to experiment design, every protocol can be broken down into individual actions. In the case of the quantum communication protocol, these actions are, e.g: apply $T$-gate to the second qubit, apply $H$-gate to the first qubit, send the third qubit to the location $B$, and measure the first qubit in the $Z$-basis. Because of the combinatorial nature of the design problem, the number of possible protocols grows exponentially with the number of available actions. For that reason, a bruteforce search of a solution is impossible for an estimated number of possible states of a quantum communication environment $0.6\times 10^{12}$~\cite{wallnofer2020machine}.

\begin{figure}[t!] 
  \centering
  \includegraphics[width=0.8\textwidth]{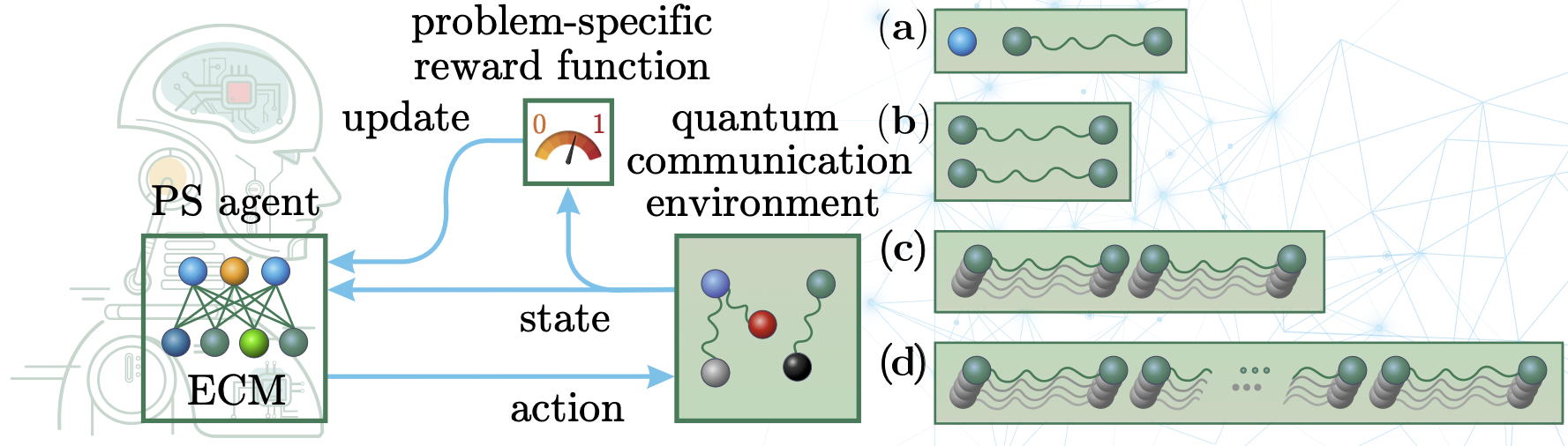}
  \caption{A reinforcement learning algorithm that designs a long-distance quantum communication protocol. The algorithm is based on projective simulation with episodic and compositional memory (ECM). Given the state of the quantum communication environment, the algorithm chooses how to modify this state by acting on the environment. The goals of the PS agent are: (a) teleportation protocol (b) entanglement purification protocol (c) quantum repeater protocol (d) quantum repeater protocol for long-distance quantum communication.}
  \label{fig:MLforQComms}
\end{figure}

A reinforcement learning approach to quantum protocol design, first proposed in Ref.~\cite{wallnofer2020machine}, is shown to be applicable to a variety of quantum communication tasks: quantum teleportation, entanglement purification, and a quantum repeater. The scheme of the learning setting is shown in Fig.~\ref{fig:MLforQComms}. The agent perceives the quantum environment state, and chooses an action based on the projective simulation deliberation process. The projective simulation network used in this work is similar to the one in Fig.~\ref{fig:MLforQExp}(b), with addition of hierarchical skill acquisition. This skill is of particular importance in the long-distance quantum communication setting, which has to include multiple repeater schemes.

With the help of projective simulation, it was demonstrated that reinforcement learning can play a helpful assisting role in designing quantum communication protocols. It is shown that the use of ML in the protocol design is not limited to rediscovering existing protocols. The agent finds new protocols that are better than existing protocols in case optimal situations lack certain symmetries assumed by the known basic protocols\footnote{Ref.~\cite{cao2022quantum} uses reinforcement learning (DPPO) to design the optimal path of quantum imaginary time evolution, which can always achieve an efficient find an efficient quantum circuit.}.

\subsection{Machine learning in random walks problems}
\label{sub-walk}

Random walks paradigm plays important role in many scientific fields related to the transfer of charge,
energy, or information transport~\cite{Giese,Lee,Farhi,kac1947random,Szameit2018}. Random (classical) walks on graphs represent indispensable tool for many subroutines in computational algorithms~\cite{rajeev1995randomized, grady2006random, gkantsidis2006random}. Quantum walks (QW) represent generalization of classical walks to quantum domain and use quantum particle instead of classical one ~\cite{aharonov1993qw, Kempe}. Resulting quantum interference pattern which governs QW physics fundamentally differs form the classical one \cite{Bromberg}. For quantum information science crucially important that quantum particle exhibits quantum parallelism which appear as a result of various paths interference and entanglement. 
It was shown that quantum particle propagates quadratically faster than classical one on certain graphs which are line ~\cite{ambainis2001one}, cycle ~\cite{solenov2006continuous,melnikov2016quantum}, hypercube ~\cite{kempe2005discrete, Krovi}, and glued trees graphs~\cite{Childs}, respectively. It is expected that algorithms based on QW should demonstrate quadratic speedup that is $O(\sqrt{N})$. Such parallelism may be useful for quantum information processing and quantum algorithms purposes~\cite{Ambainis, Childs,venegas2008quantum}. 
It is especially important to note that QW are explored in quantum search algorithms which represent important tools for speedup of QML algorithms ~\cite{Biamonte, portugal2013quantum, paparo2014quantum, chakraborty1016spatial}.

Noteworthy, QW speedup demonstration with arbitrary graphs represents open problem~\cite{melnikov2019hitting}. Standard approach would be to simulate quantum and classical dynamics on given graph, which provides an answer in which case a particle would arrive a target vertex faster. However, this approach may be difficult (and costly) to use in computations for the graphs possessing large number of vertices; the propagation time scales polynomially in the size of the graph. Second, we usually interested in a set of graphs for which obtained results of the simulations cannot reveal some general features of quantum advantage. 

In a number of works we attacked this problem by means of ML approach~\cite{melnikov2019predicting, melnikov2020machinetransfer, melnikov2022}. We explore a supervised learning approach to predict a quantum speedup just by looking at a graph. In particular, we designed a classical-quantum convolutional neural network (CQCNN) that learns from graph samples to recognize the quantum speedup of random walks.

The basic concept of CQCNN that we use in Ref.~\cite{melnikov2019predicting, melnikov2020machinetransfer, melnikov2022} is shown in Fig.~\ref{fig:graph} and Fig.~\ref{fig:CQCNN}, respectively. In particular, we examined in Ref.~\cite{melnikov2019predicting, melnikov2020machinetransfer,melnikov2020training,melnikov2020deep, melnikov2022} continuous-time random walks and suppose that classical random walk representing stochastic (Markovian) process defined on a connected graph. It starts at the time $t=0$ from initial node $i$ and hits target vertex $t$. Unlike the classical case, a quantum particle due to interference phenomenon will be ``smeared" across all vertices of the graphs. Thus, in the quantum case we propose an additional (sink) vertex $s$ that is connected to the target vertex and provides localization of the quantum particle due to energy relaxation from $t$ to $s$ vertices which happen with the rate $\gamma$. In other words, the quantum particle may be permanently monitored in the sink vertex. Mathematically graph is characterized by it weighted adjacency matrix $A$ that is relevant to Hamiltonian $\mathcal{H}=\hbar A$. Notice, for chiral QW time asymmetry may be obtained by using complex-valued adjacency matrix elements~\cite{melnikov2022}. We characterize quantum transport by means of Gorini–Kossakowski–Sudarshan–Lindblad equation that look like (cf.~\cite{Manzano}):

\begin{figure}[t!] 
  \centering
  \includegraphics[width=0.8\textwidth]{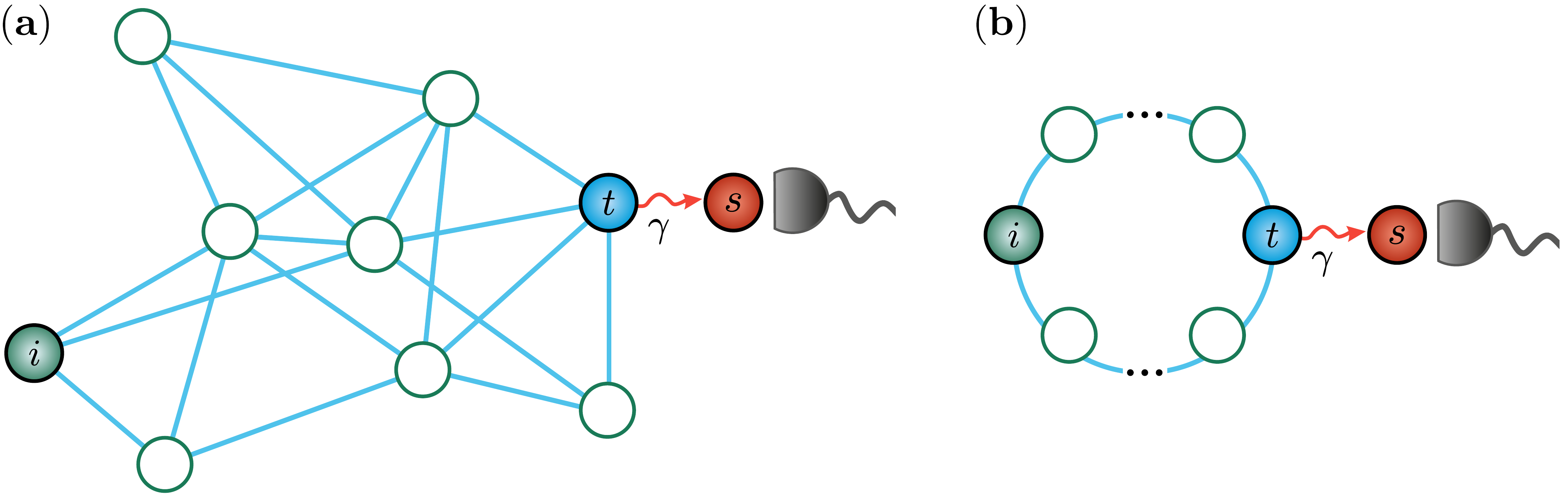}
  \caption{A schematic representation of considered in Refs.~\cite{melnikov2019predicting, melnikov2020machinetransfer} random walks on  (a) connected random  graph, (b) cycle graph. The  labels ($i$), and ($t$)  specify  initial and target vertices, respectively; $s$ is a sink vertex which is require to localize and detect quantum particle. The $\gamma$ is  coupling parameter between target and sink vertices, respectively.}
  \label{fig:graph}
\end{figure}

\begin{align}\label{qwdynamics}
  &\frac{\mathrm{d}\rho(t)}{\mathrm{d}t} = - \frac{i}{\hbar}(1-p)\left[\mathcal{H},\rho(t)\right]\nonumber
   + p\sum_{mk}\left( L_{mk}\rho(t) L_{mk}^\dagger - \frac{1}{2}\left\{L_{mk}^\dagger L_{mk},\rho(t)\right\}\right)\nonumber\\
   &+ \gamma\left(L_s\rho(t)L_s^\dagger -\frac{1}{2}\left\{ L_s^\dagger L_s, \rho(t)\right\}\right), 
   \end{align}
where $\rho(t)$ is time dependent density operator, $L_{mk}=T_{mk}\ket{m}\bra{k}$ and $L_{s}=\ket{s}\bra{t}$ operators characterize transitions from vertices $k$ to $m$ and from $t$ (target) to $s$ (sink), respectively; $\gamma$ is coupling parameter between target and sink vertices. The parameter $p$ lies $0\leq p\leq 1$ condition and determines the decoherence; the value $p=0$ is relevant to purely quantum transport, while $p=1$ determines completely classical random walks.

Solution of Eq.~\eqref{qwdynamics} specifies quantum probability $\mathrm{P^q}(t)\equiv\rho_{(n+1)(n+1)}(t)$ ($n$ is the total number of vertices), which is relevant to QW on a chosen graph. The classical random walk may be established by the probability distribution
\begin{equation}\label{rwdynamics}
    \mathrm{P(t)} = \mathrm{e}^{-It}\mathrm{e}^{Tt}\mathrm{P(0)},
\end{equation}
where $\mathrm{P}(t)$ is a vector of probabilities $\mathrm{P}_v(t)$ of detecting a classical particle in vertices $v\in \mathcal{V}$ of the graph; $I$ is the identity matrix of size $n\times n$. The transition matrix $T$ is a matrix of probabilities $T_{vu}$ for a particle to jump from $u$ to $v$. In this case the sink vertex is not needed and we can assume $\gamma=0$. We are interested in the probability of finding a particle in the target (or, in the sink) vertex which is described by solutions of Eqs.~(\ref{qwdynamics}) and (\ref{rwdynamics}).

 \begin{figure*}[!ht]
	\centering
	\includegraphics[width=0.8\textwidth]{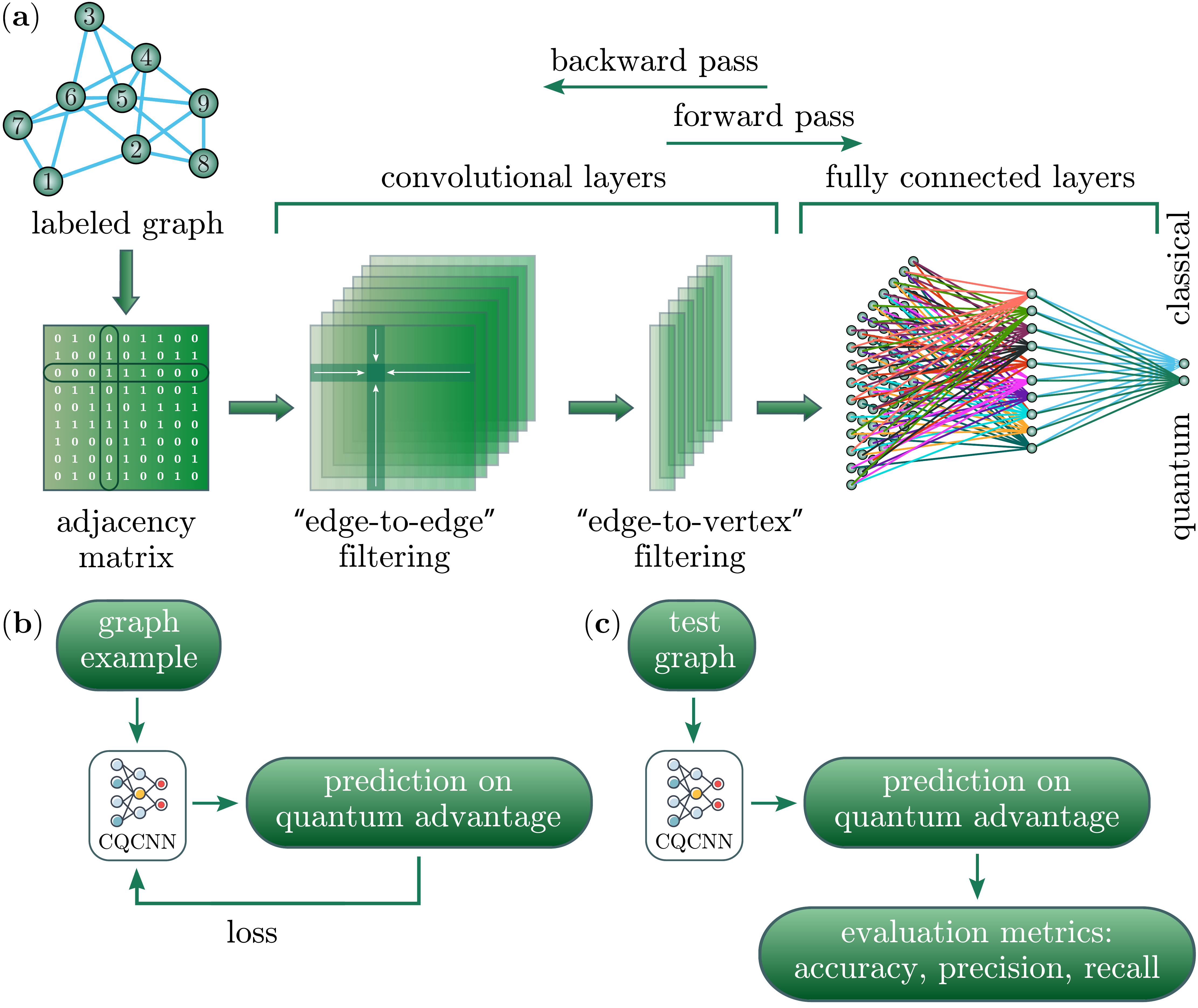}
	\caption{Schematic representation of CQCNN approach which is used for predicting the quantum speedup on the graphs represented in Fig. \ref{fig:graph}. (a) - scheme of the CQCNN architecture. The neural network takes a labeled graph in form of an adjacency matrix $A$ as an input. The $A$ then processed by convolutional layers with filters of graph-specific ``edge-to-edge" and ``edge-to-vertex", respectively. These filters act as functions of a weighted total number of neighboring vertices of each vertex. The convolutional layers are connected with fully-connected layers which classify the input graph.  Data and error propagation are shown with arrows. (b) and (c) demonstrates  processes of CQCNN training and testing, respectively.}
	\label{fig:CQCNN}
\end{figure*}

Then, one possible to compare $\mathrm{P^q}(t)\equiv\rho_{(n+1)(n+1)}(t)$ and $\mathrm{P^c}(t)\equiv\mathrm{P}_n(t)$ against $\mathrm{P}_{th}=1/\log n$ that determines threshold value of probability for a given graph. If this probability is larger than $\mathrm{P}_{th}$, we can conclude that the particle occurs at the target. The time at which one of inequalities $\mathrm{P^q}(t)>\mathrm{P}_{th}$, $\mathrm{P^c}(t)>\mathrm{P}_{th}$ fulfilled, is called the hitting time for quantum or classical particle, respectively. Hence, by comparing the solutions to Eqs.~(\ref{qwdynamics}) and~(\ref{rwdynamics}), we can define the particle transfer efficiency: it is $1$ if the quantum particle reached the target first, and $0$ otherwise.

In Fig.~\ref{fig:CQCNN} we schematically summarize proposed CQCNN approach for detection of QW speedup. In order, the architecture of CQCNN is shown in Fig.\ref{fig:CQCNN}(a). It consists of a two-dimensional input layer that takes one graph represented by an adjacency matrix $A$. This layer is connected to several convolutional layers, the number of which depends on the number of vertices $n$ of the input graph. The number of layers is the same for all graph sizes. CQCNN has a layout with convolutional and fully connected layers, and two output neurons that specify two possible output classes. The convolutional layers are used to extract features from graphs, and decrease the dimensionality of the input. 

Empirically we find out that relevant features are in the rows and columns of adjacency matrices. The first convolutional layer comprises six filters (or, feature detectors) which define three different ways of processing the input graph. These three ways are marked by green, red, blue colors in Fig.~\ref{fig:CQCNN}(a), respectively. The constructed filters are form ‘crosses’ which are shown in Fig.~\ref{fig:CQCNN}(a) and capture a weighted sum of column and row elements. These filters act as functions of a weighted total number of neighboring vertices of each vertex. Thus, the cross ‘edge-to-edge’ and ‘edge-to-vertex’ filters crucially important in designed CQCNN; they are capable for prediction of the quantum advantage by QW.
 
Fig.~\ref{fig:CQCNN}(b) shows schematically training procedure by using some graphs samples which are established by adjacency matrices $A$ as an input. CQCNN made prediction at the output determining classical or quantum classes depending on the values of two output neurons. The predicted class is determined by means of the index of a neuron with the largest output value $\mathrm {class} = \mathrm {argmax}_m y(m)$. Having a correct label, the loss value is computed.

The filters that we constructed in CQCNN play an essential role in the success of learning. CQCNN learns by stochastic gradient descent algorithm that takes the cross entropy loss function. The loss on a test example $i$ is defined relative to the correct class $\mathrm{class}_i$ (classical or quantum, $0$ or $1$) of this example:
\begin{equation}\label{lossFunction}
\mathrm{loss}_i = - \kappa(\mathrm{class}_i)\log \left(\frac{\mathrm{e}^{x(\mathrm{class}_i)}}{\mathrm{e}^{x(0)}+ \mathrm{e}^{x(1)}}\right),
\end{equation}
where we defined the values of the output neurons as $x(0)$ and $x(1)$; $\kappa(\mathrm{class}_i)$ is the total fraction of examples from this class in the dataset. As we have shown in Ref.~\cite{melnikov2019predicting} CQCNN constructed a function that generalizes over seen graphs to unseen graphs, as the classification accuracy (which may be defined as the fraction of correct predictions) goes up.
 
CQCNN testing procedure is not principally different from the training process how it is seen in Fig.~\ref{fig:CQCNN}(c), cf. Fig.~\ref{fig:CQCNN}(b). The CQCNN does not receive any feedback on its prediction in this case and the network is not modified. 

We apply the described ML approach to different sets of graphs. In particular, to understand how our approach works in a systematic way, we first analyze the CQCNN on line graphs with up to $10$ vertices. CQCNN was trained over $2000$ epochs with a single batch of $3$ examples per epoch. 

\begin{figure*}[ht!]
	\centering
	\includegraphics[width=0.8\textwidth]{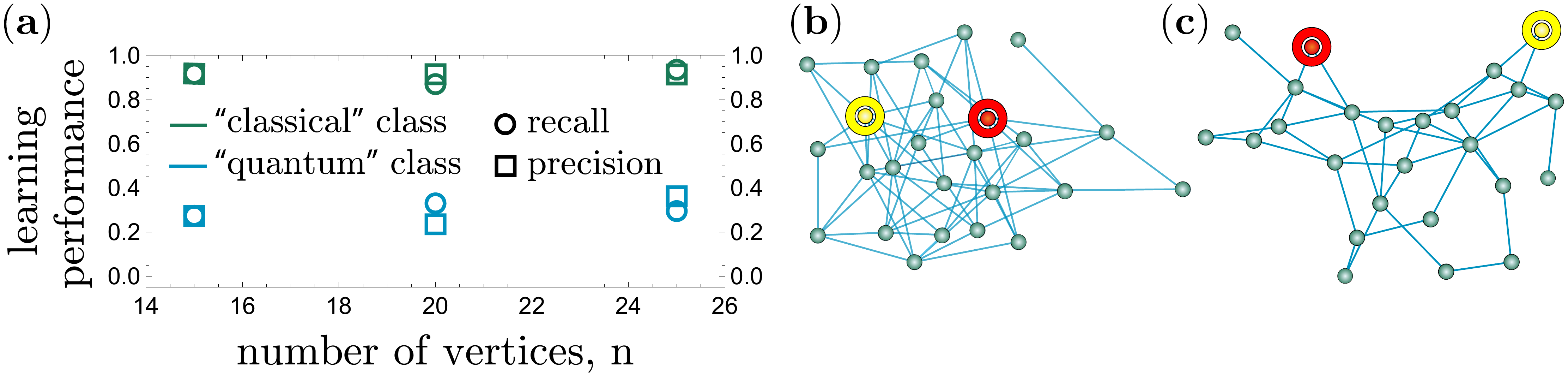}
	\caption{(a) CQCNN learning performance. Dataset consists of random graphs with $n=15, 20$ and $25$ vertices, $1000$ examples for each $n$, and the corresponding classical and quantum labels. CQCNN was simulated during $3000$ epochs, $100$ mini batches each with the batch size of $3$ examples. The neural network was tested on $1000$ random graphs for each $n$. (b), (c) establish random graph examples taken from the test set which were correctly classified by CQCNN (initial and target vertices are marked in yellow and red, respectively). The classical particle is faster on (b), whereas the quantum one is faster on graph (c).}
	\label{fig:randomLearning}
\end{figure*}

Then, we simulated CQCNN's learning process for random graphs, each sampled uniformly from the set of all possible graphs with $n$ vertices and $m$ edges. The learning performance results in the absence of decoherence ($p=0$) are shown in Fig.~\ref{fig:randomLearning} for $n=15, 20, 25$; the $m$ is chosen uniformly from $n-1$ to $(n^2-n)/2$. 
Our simulations shows that the loss after training is vanishing; it is below $3\times 10^{-3}$ for all these random graphs. 
In Fig.~\ref{fig:randomLearning}(a) we see that both recall and precision are about $90\%$ for the ~``classical'' part of the set, and is in the range of $25-35\%$ for the ``quantum'' one. n\footnote{ 
Recall quantifies the fraction of correct predictions in a particular class, whereas precision identifies the fraction of a particular class predictions that turned out to be correct.}
Thus, one can see that CQCNN helps to classify random graphs correctly much better than a random guess without performing any QW dynamics simulations. In Fig.~\ref{fig:randomLearning}(b)-(c) we represent samples of correctly classified graphs. 

\subsubsection{Quantum walks with decoherence}

\begin{figure}[t!] 
 \centering
 \includegraphics[width=0.8\textwidth]{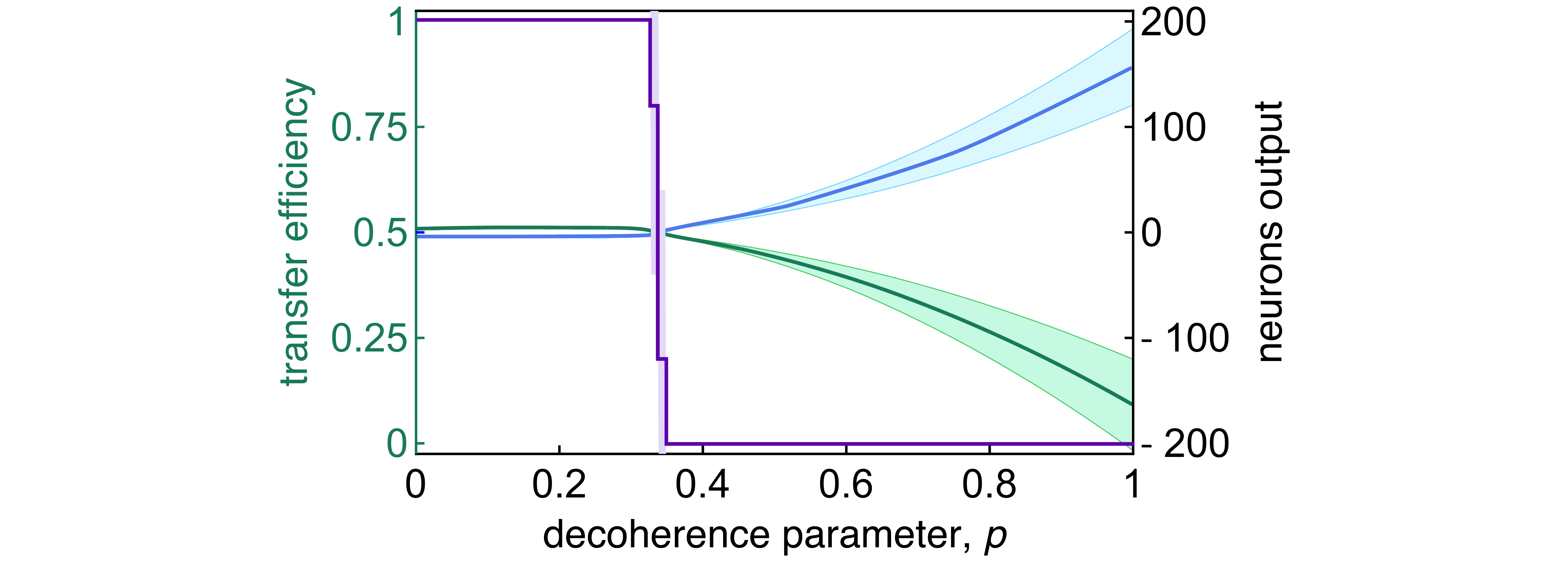}
 \caption{Prediction of transfer efficiency (violet curve) for a $6$-cycle graph versus decoherence parameter $p$. The activation values of output neurons are shown in blue and green. The results obtained by averaging of $5$ CQCNN networks. Standard deviations are marked by shaded regions.}
 \label{fig:6cycle}
\end{figure}

In the presence of decoherence, i.e. for $p>0$ physical picture is getting richer. In Fig.~\ref{fig:6cycle} we demonstrate results of QW dynamics simulation on cycle graph consisting of $6$-vertices; the efficiency of transport is measured between opposite vertices of the graph as it is shown in Fig.~\ref{fig:graph}(b). Simulations performed for $1000$ randomly sampled values of the decoherence parameter $p$ and used to train CQCNN. After the training procedure we suggest CQCNN to predict if the QW can lead to an advantage for a new given parameter $p$. In Fig.~\ref{fig:6cycle} we represent the results of the transfer efficiency predictions as a violet line. From Fig.~\ref{fig:6cycle} clearly seen that at the value of decoherence parameter $p\simeq0.34$ abrupt crossover from quantum ($p\simeq1$) to classical ($p\simeq0$) regime transport occurs. Thus, one possible to expect QW advantage in transport in domain of $p<0.34$. Physically, such an crossover may be relevant to quantum tunneling features in the presence of dissipation, cf. ~\cite{larkin1983quantum,grabert1984quantum}. Notice that the parameter $p$ is temperature-dependent in general, cf. ~\cite{Alodjants2017}. In this case we can recognize the established crossover as a (second-order) phase transition from quantum to classical (thermal activation) regime that happens for a graph at some finite temperature.

The Fig.~\ref{fig:6cycle} demonstrates predictions of CQCNN which are based on the learned values of the output neurons; they are shown in Fig.~\ref{fig:6cycle} as a classical (blue) and quantum (green) classes, respectively. CQCNN made decision about the class by using the maximum value of the output neurons activation. From Fig.~\ref{fig:6cycle} it is clearly seen that the ``vote'' for the quantum class grows up to the maximum value $p\simeq 0.2$, which corresponds to the highest confidence for the quantum class. Simultaneously, the confidence in the classical class grows with increasing of decoherence parameter $p$. Separation between classes become more evident after the crossover point of $p\simeq 0.34$.

Thus, obtained results play significant role in creation of soft- and hardware systems which are based on the graph approach at their basis. CQCNN that we proposed here allows to find out which graphs, and under which conditions on decoherence, can provide a quantum advantage. This is especially relevant to NISQ era quantum devises development. 

\subsection{Machine learning in quantum tomography}
\label{sub-tomography}
With the capability to find the best fit to arbitrarily complicated data patterns with a limited number of parameters available, machine learning has provided a powerful approach for quantum tomography. Here, quantum tomography or quantum state tomography (QST) refers to the reconstruction  about a quantum state with its comprehensive information by using measurements on an ensemble of identical quantum states~\cite{Hradil, PhysRevA.63.040303, PhysRevA.64.052312, Blume_Kohout_2010, PhysRevA.68.012305, zbMATH01476719, QST}. However, the exponential growth in bases for a Hilbert space of $N$-qubit states implies that exact tomography techniques require exponential measurements and/or calculations. In order to leverage the full power of quantum states and related quantum processes, a well characterization and validation of large quantum systems is demanded and remains important challenge~\cite{PRXQuantum.2.040201}. 

Traditionally,  by estimating the closest probability distribution to the data for any arbitrary quantum states, the maximum likelihood estimation (MLE) method is used in quantum tomography~\cite{MLE, PhysRevLett.108.070502}. 
However, MLE method requires exponential-in-$N$ amount of data as well as an exponential-in-$N$ time for processing. 
Albeit dealing with Gaussian quantum states, unavoidable coupling from the noisy environment makes a precise characterization on the quantum features in a large Hilbert space almost unattackable.
Moreover, MLE also suffers from the over-fitting problem when the number of bases grows. 
To make QST more accessible, several alternative algorithms are proposed by assuming some physical restrictions imposed upon the state in question, such as the adaptive quantum tomography~\cite{Granade_2017},  permutationally invariant tomography~\cite{PhysRevLett.105.250403}, quantum compressed sensing~\cite{Flammia_2012, PhysRevLett.105.150401, PhysRevLett.106.100401, Riofrio2017}, tensor networks~\cite{tensor-1, tensor-2}, generative models~\cite{carrasquilla2019reconstructing}, feed-forward neural networks~\cite{npj-local}, and variational autoencoders~\cite{Rocchetto2018}.

\begin{figure*}[t!]
	\centering
	\includegraphics[width=0.8\textwidth]{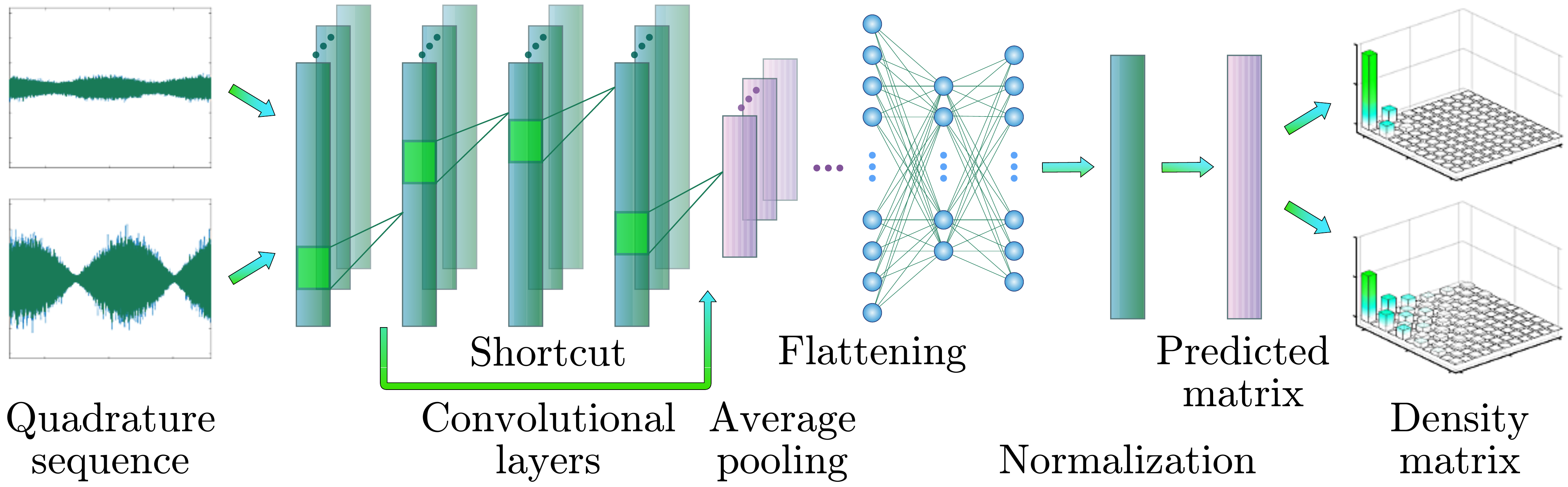}
	\caption{Schematic of machine learning enhanced quantum state tomography with convolutional neural network (CNN). Here, the noisy data of quadrature sequence obtained by quantum homodyne tomography in a single-scan are fed to the convolutional layers, with the shortcut and average pooling in the architecture. Then, after flattening and normalization, the predicted matrices are inverted to reconstruct the density matrices in truncation.}
	\label{fig:CNN-QST}
\end{figure*}

To reduce the over-fitting problem in MLE, the restricted Boltzmann machine (RBM)~\cite{RBM, 10.1162/neco.2008.04-07-510, s15516709cog0901, PhysRevB.100.195125}  has provided a powerful solution in QST. With the help of two layers of stochastic binary units, a visible layer and a hidden layer, the RBM acts as a universal function approximator. For qubits on an IBM Q quantum computer, quantum state reconstructions via ML were demonstrated with four qubits~\cite{9521839}. For continuous variables, the convolutional neural network (CNN) has been experimentally implemented with the quantum homodyne tomography for continuous variables~\cite{PhysRevA.102.022412, palmieri2020experimental, PhysRevLett.128.073604}.

As illustrated in Fig.~\ref{fig:CNN-QST}, the time sequence data obtained in the optical homodyne measurements share the similarity to the voice (sound) pattern recognition~\cite{10.1145/3065386,7780459}. Here, the noisy data of quadrature sequence are fed into a CNN, composited with $30$ convolutional layers in total. In applying CNN, we take the advantage of  good generalizability to extract the resulting density matrix from the time series data~\cite{10.1145/3065386}. In our deep CNN, there are  four convolution blocks used, each  containing $1$ to $9$ convolution layers (filters) in different sizes. Five shortcuts are also introduced among the convolution blocks, in order to tackle the gradient vanishing problem. Instead of max-pooling, average pooling is applied to produce higher fidelity results, as all the tomography data should be equally weighted. Finally, after flattening two fully connected layers  and normalization, the predicted matrices are inverted to reconstruct the density matrices in truncation.

Here, the loss function we want to minimize is the mean squared error (MSE); while the optimizer used for training is Adam. We take the batch size as $32$ in the training process. By this setting, the network is trained with $70$ epochs to decrease the loss (MSE) up to $5\times 10^{-6}$. Practically, instead of an infinite sum on the photon number basis, we keep the sum in the probability up to $0.9999$ by truncating the photon number. Here, the resulting density matrix is represented in photon number basis, which is  truncated to $35\times 35$ by considering the maximum anti-squeezing level up to $20$~dB.

As to avoid non-physical states, we impose the positive semi-definite constraint into the predicted density matrix. Here, an auxiliary (lower triangular) matrix is introduced before generating the predicted factorized density matrix  through the Cholesky decomposition. During  the training process, the normalization also ensures  the trace of the output density matrix is kept as $1$. More than a million data sets are fed into our CNN machine with a variety of squeezed ($\rho^{sq}$), squeezed thermal ($\rho^{sq}_{th}$), and thermal states ($\rho_{th}$) in different squeezing levels, quadrature angles, and reservoir temperatures, i.e.,
\begin{eqnarray}
&&\rho^{sq} =\hat{S}\rho_{0}\hat{S}^\dag;\\
&& \rho^{sq}_{th} =\hat{S}\rho_{th}\hat{S}^\dag.
\end{eqnarray}
Here, $\rho_0 = |0\rangle\langle 0|$ with the vacuum state $|0\rangle$, $\rho_{th} = \sum_{n} P(n) \, |n\rangle\langle n|$ with the probability distribution function $P(n) = \frac{1}{\bar{n}+1}(\frac{\bar{n}}{\bar{n}+1})^n$, defined with the mean-photon number $\bar{n} =  \frac{1}{\text{exp}[\hbar \omega/k_B T]-1}$ at a fitting temperature $T$, and $\hat{S} (\xi) = \text{exp}[\frac{1}{2} \xi^\ast \hat{a}^2 - \frac{1}{2} \xi \hat{a}^{\dag 2}]$ denotes the squeezing operator, with the squeezing parameter $\xi \equiv r\, \text{exp}(i \phi)$ characterized by the squeezing factor $r$ and the squeezing angle $\phi$. All the training is carried out with the Python package {\it tensorflow.keras} performed in GPU  (Nvidia Titan RTX).

The validation of ML-enhanced QST is verified with simulated data set, through the average fidelity obtained by MLE and CNN by calculating the purity of  quantum state, i.e., $\text{purity} \equiv \text{tr}(\rho^2)$. Compared with the time-consuming MLE method, ML-enhanced QST keeps the fidelity up to $0.99$ even taking $20$~dB anti-squeezing level into consideration. With prior knowledge in squeezed states, such a supervised CNN machine can be trained in a very short time (typically in less than one hour), enabling us to build a specific machine learning for certain kinds of problem. When well-trained, an average time of about $38.1$ milliseconds (by averaging $100$ times) costs in a standard GPU server. One unique advantage of ML-enhanced QST is that we can  precisely identify the pure squeezed and noisy parts in extracting the degradation information. By directly applying the singular value decomposition to the predicted density matrix, i.e.,  $\rho = \sigma_1\, \rho^{sq}+ c_1\, \rho^{sq}_{th} + d_1\, \rho_{th}$, all the weighting ratio about the ideal (pure) squeezed state, the squeezed thermal state, and thermal state can be obtained. With this identification, one should be able to suppress and/or control the degradation at higher squeezing levels, which should be immediately applied to the applications for gravitational wave detectors and quantum photonic computing.
 
Toward a real-time QST to give physical descriptions of every feature observed in the quantum noise, a characteristic model to directly predict physical parameters in such a CNN configuration is also demonstrated~\cite{sym14050874}. Without dealing with a density matrix in a higher dimensional Hilbert space, the predicted physical parameters obtained by the characteristic model are as good as those generated by a reconstruction model. One of the most promising advantages for ML in QST is that only fewer measurement settings are 
needed~\cite{PhysRevLett.127.130503}. Even with incomplete projective measurements, the resilience of ML-based  quantum state estimation techniques was demonstrated  from partial measurement results~\cite{Danaci_2021}. Furthermore, such a high-performance, lightweight, and easy-to-install supervised characteristic model-based ML-QST can be easily installed on edge devices such as FPGA as an in-line diagnostic toolbox for all possible applications with squeezed states. 

In addition to the squeezed states illustrated here, similar machine learning concepts can be readily applied to a specific family of continuous variables, such as non-Gaussian states. 
Of course, different learning (adaptation) processes should be applied in dealing with single-photon states, Schrödinger’s cat states~\cite{RevModPhys.85.1083, RevModPhys.85.1103}, and Gottesman-Kitaev-Preskill states for quantum error code corrections~\cite{GKP}. 
Alternatively, it is possible to use less training data with a better kernel developed in machine learning, such as the reinforce learning, generative adversarial network, and the deep reinforcement learning used in the optimization problems~\cite{torlai2019integrating, PhysRevLett.127.140502, PhysRevResearch.3.033278, Quek2021, Nakaji2021, Haug_2020, PhysRevA.106.012409, Cha_2021}.
Even without any prior information, informational completeness certification net (ICCNet), along with a fidelity prediction net (FidNet), have also been carried out to uniquely reconstruct any given quantum state~\cite{Teo_2021}.

Applications of these data-driven learning and/or adaptation ML are not limited to quantum state tomography. Identification and estimation of quantum process tomography, Hamiltonian tomography, and quantum channel tomography, as well as quantum phase estimation, are also in progresses~\cite{PhysRevA.90.010103, PhysRevA.88.022120, PhysRevLett.93.080502, PhysRevA.104.062404}.
Moreover, ML in quantum tomography can  be used for the quantum state preparation, for the general single-preparation quantum information processing (SIPQIP) framework~\cite{9583935}.

\subsection{Photonic quantum computing}

In addition to classical information processing, photonic quantum computing is also one of the possible technologies to demonstrate quantum advantage~\cite{Arute, Zhong2020, PhysRevLett.127.180502}, i.e., in which a quantum system has been shown to outperform a classical one on some well-defined information processing task. 
Even though the computational task on the implementation of photonic Boson sampling  is non-universal~\cite{10.1145/1993636.1993682}, meaning that it cannot perform arbitrary quantum operations,  whether any useful applications exist within the heavily restricted space of non-universal photonic systems is an open question.
Nevertheless, the advantages of photonics as a quantum technologies platform compared to other platforms are a high degree of integration with mature classical photonic technologies, and the fact that the photonic circuits involved can be operated at room temperature~\cite{Wang2020, Elshaari2020}.

Reviews on the recent advances of machine learning, in particular deep learning, for the photonic structure design and optical data analysis, as well as the challenges and perspectives, can be found in Refs.~\cite{Ma-Rev, Duan-Rev, 9064516, https://doi.org/10.1002/adma.201901111}.
Inverse designs and optimization on the photonic crystals, plasmonic nano-structures, programmable meta-materials, and meta-surfaces have been actively explored for high-speed optical communication and computing, ultrasensitive biochemical detection, efficient solar energy harvesting, and super-resolution imaging~\cite{quantum1010011}. By utilizing adaptive linear optics, quantum machine learning is proposed for performing quantum variational classification and quantum kernel estimation~\cite{Chabaud2021quantummachine}.

Towards the goal of realization of on-chip quantum photonics, silicon-based materials have been actively explored due to their compatibility with conventional CMOS fabrication processes. 
Instead of using photons or entangled photons pairs from optical parametric process, which suffers from the low probabilities and rare success events in post-selection, quantum optical microresonator on a chip employs the Kerr nonlinearity has the advantage over the photonic qubit-based approach~\cite{Reimer2016, Kues2017,  Kues2019, Yang2021, Guidry2022}. Unlike all gate-based quantum computing, which are based on single-qubit and entangling gates, photonic quantum computing does not rely on any physical gates during the computation, but on the  preparation of quantum modes, e.g., cluster states,  with the time-domain and/or frequency-domain multiplexings~\cite{Warit2019, Larsen2019}.

Regarding the integrated quantum photonics, as well as the progress of the hybrid quantum devices, machine-learning methodology has also been widely applied in order to offer an efficient means to design photonic structures, as well as  to tailor light–matter interaction~\cite{Kudyshev-ACS}. For the emerging field of machine learning quantum photonics, we need not ML for automated design of quantum optical experiments~\cite{PhysRevX.11.031044}, such as the quantum walk with graph states illustrated in Sec.~\ref{sub-walk}, but also quantum technologies enhanced by machine learning, such as ML in quantum tomography illustrated in Sec.~\ref{sub-tomography} to have ML-assisted quantum measurements. Along this direction, quantum optical neural network (QONN) was introduced to leverage advances in integrated quantum photonics~\cite{Steinbrecher2019}. With the newly developed protocols such as quantum optical state compression for quantum networking and black-box quantum simulation, many thousands of optoelectronic components should be integrated in  monolithically integrated circuits. It is expected to see the combination of quantum measurements, quantum metrology, and optimization of on-chip and free-space meta-devices for photonic quantum computing as a promising route for automatization of quantum experiments.

\section{Conclusions}

\begin{figure}
    \centering
    \includegraphics[width=0.8\textwidth]{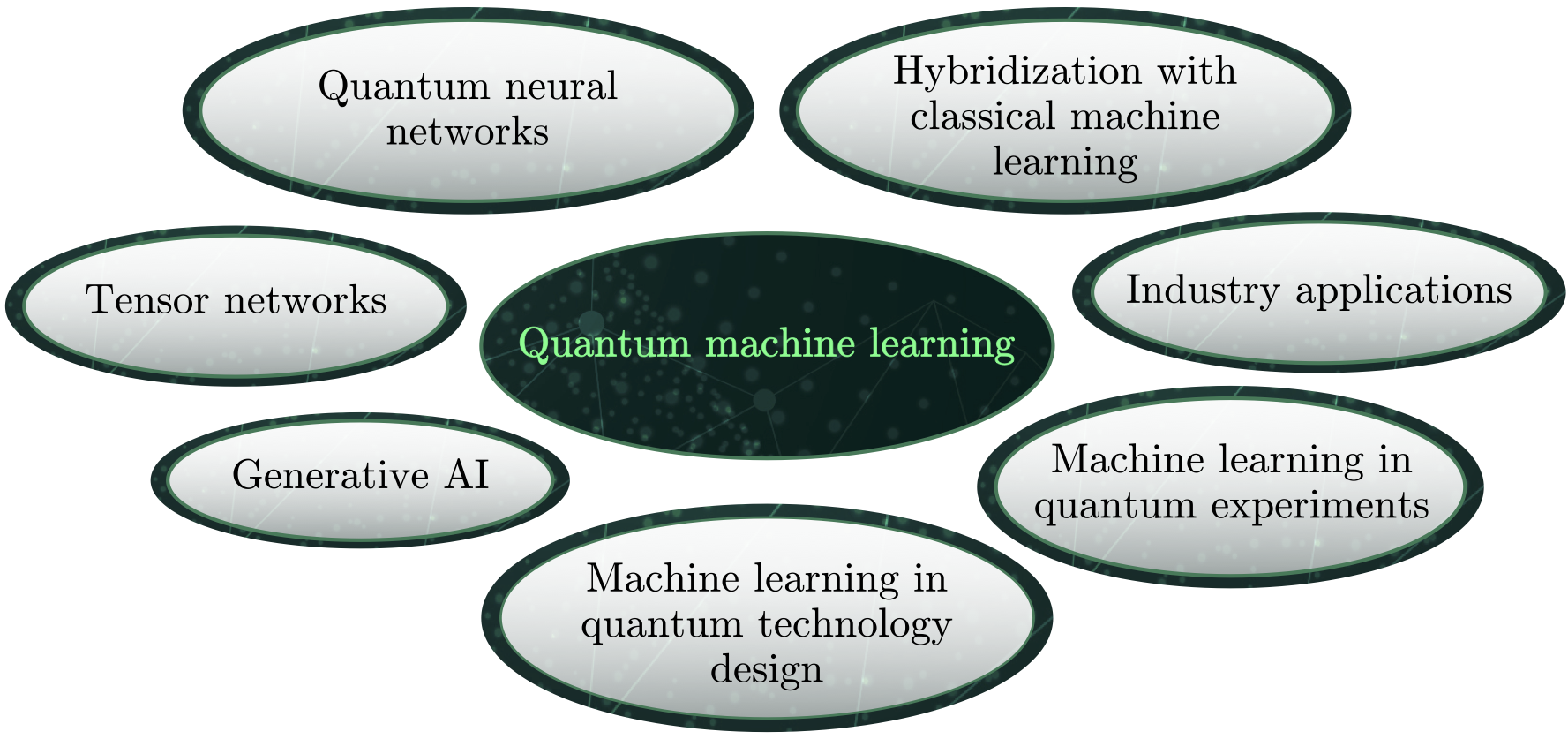}
    \caption{Quantum machine learning ecosystem for the next decade. Please check the references in the published paper version: Alexey Melnikov, Mohammad Kordzanganeh, Alexander Alodjants, Ray-Kuang Lee (2023) Quantum machine learning: from physics to software engineering, Advances in Physics: X, 8:1, DOI: \href{https://doi.org/10.1080/23746149.2023.2165452}{10.1080/23746149.2023.2165452}}
    \label{fig:15}
\end{figure}

Artificial intelligence and machine learning are currently key factors in the progress of modern society in the world. Now it is quite difficult to find an area of our life where achievements in the field of artificial intelligence would not be used. However, this success is largely ensured by the development of classical information technologies in terms of hardware, which possess  natural limitations. The creation of quantum computers and quantum networks can bypass these limitations in different fields. Quantum machine learning is a rapidly developing new field  of research at the border  of artificial intelligence, quantum information science  and quantum technology.

The ecosystem of QML, which has developed to date and will be in demand for the next decade, is depicted in Fig.\ref{fig:15}. 
Here we proceed from the fact that in the near future, the main role in our daily life will be played by various (tensor) network systems of transmission, processing, and intelligent recognition of large amounts of information. In this regard, we are confident that practically significant quantum computers will be embedded  into large distributed intelligent systems  environment  that will surround us everywhere.

In this review, we outlined the hot topics of interplay between promising artificial intelligence methods and modern quantum computing. These topics are predominantly associated with the limited capabilities of NISQ era quantum computers and use variety of variational algorithms such as variational eigensolver and quantum approximate optimisation algorithm, respectively.  A special place in our review is given to  so-called quantum neural networks which represent  new QML models whose parameters are updated classically and may be used within  quantum-classical training algorithms for variational circuits. In this sense we discussed  promising hybrid information processing methods that use both classical ML algorithms and quantum devices. The training procedure proposes data providing to the quantum model, calculating an objective (loss) value, and then adapting the QNN parameters. Thus, the whole procedure represents a hybrid quantum-classical algorithm. In this work we discuss another possible application for hybrid computation, that may use classical-quantum convolutional neural network designed for resolving speedup of random walks on chosen graphs. Our  approach is based on training CQCNN, which learns to extract feature vectors from graph adjacency matrices combined with a decoherence parameter. We have shown that even without any decoherence in generally the speedup of random walks essentially depends on topological properties of the graph, i.e. on  adjacency matrix peculiarities. Our findings  open new perspectives in quantum-classical algorithms which explore random walks as subroutines.

\section{Acknowledgements}
A.P.A acknowledges  support from the Goszadanie No. 2019-1339  project of Ministry of Science and Higher Education of Russian Federation.  R.-K.L. is partially supported by the National Science and Technology Council of Taiwan (No. 110-2123-M-007-002).

\bibliography{qml}

\end{document}